\documentclass[iop,apj]{emulateapj}
\usepackage{CJK,amsmath}
\pdfoutput=1

\renewcommand{\vec}[1]{\boldsymbol{#1}}

\newcommand{\pder}[2]{\frac{\partial #1}{\partial #2}}
\newcommand{\tder}[2]{\frac{\mathrm{d}#1}{\mathrm{d}#2}}
\newcommand{\tdder}[2]{\frac{\mathrm{d}^2#1}{\mathrm{d}#2^2}}
\newcommand{\unitvec}{\hat{\vec{e}}}
\newcommand{\epsj}{\tilde{\epsilon}_{kj}}
\newcommand{\eps}{\epsilon_{tot}}
\newcommand{\ope}{1 + \eps}
\newcommand{\pope}{\left(\ope\right)}
\newcommand{\ot}{\omega t}
\newcommand{\ucm}{\bar{\vec{u}}}
\newcommand{\twodim}[2]{#1$\times$#2}
\newcommand{\threedim}[3]{#1$\times$#2$\times$#3}

\shorttitle{Particle-Gas Systems with Stiff Mutual Drag}
\shortauthors{Yang \& Johansen}

\begin{document}
\begin{CJK*}{UTF8}{bkai}

\title{Integration of Particle-Gas Systems with Stiff Mutual Drag Interaction}
       
\author{Chao-Chin Yang (楊朝欽) and Anders Johansen}
\affil{Lund Observatory, Department of Astronomy and Theoretical Physics,
       Lund University, Box~43, SE-221\,00 Lund, Sweden}
\email{ccyang@astro.lu.se; anders@astro.lu.se}

\begin{abstract}
Numerical simulation of numerous mm/cm-sized particles embedded in a gaseous disk has become an important tool in the study of planet formation and in understanding the dust distribution in observed protoplanetary disks.
However, the mutual drag force between the gas and the particles can become so stiff, particularly because of small particles and/or strong local solid concentration, that an explicit integration of this system is computationally formidable.
In this work, we consider the integration of the mutual drag force in a system of Eulerian gas and Lagrangian solid particles.
Despite the entanglement between the gas and the particles under the particle-mesh construct, we are able to devise a numerical algorithm that effectively decomposes the globally coupled system of equations for the mutual drag force and makes it possible to integrate this system on a cell-by-cell basis, which considerably reduces the computational task required.
We use an analytical solution for the temporal evolution of each cell to relieve the time-step constraint posed by the mutual drag force as well as to achieve the highest degree of accuracy.
To validate our algorithm, we use an extensive suite of benchmarks with known solutions in one, two, and three dimensions, including the linear growth and the nonlinear saturation of the streaming instability.
We demonstrate numerical convergence and satisfactory consistency in all cases.  Our algorithm can for example be applied to model the evolution of the streaming instability with mm/cm-sized pebbles at high mass loading, which has important consequences for the formation scenarios of planetesimals.
\end{abstract}


\section{INTRODUCTION} \label{S:intro}

Understanding dust dynamics in protoplanetary disks is an important topic in the study of planet formation, both theoretically and observationally.
On one hand, planetary cores must be built up from micron-sized interstellar dust grains, which need to gradually grow in size within a gaseous environment \citep{vS69}.
Especially challenging is how mm/cm-sized particles can effectively sediment towards the mid-plane of a protoplanetary disk and coalesce to form km-sized planetesimals \citep[][and references therein]{JB14}.
Thanks to the advance in telescopes, on the other hand, it has become possible for observations to resolve the distribution of mm/cm-sized particles in nearby protoplanetary disks, from their thermal emission in sub-millimeter and radio wavelengths or polarized scattered light in the near infrared.
For instance, \cite{MDB13} used the Atacama Large Millimeter/centimeter Array (ALMA) to locate large-scale, lopsided concentration of mm-sized pebbles in the transition disk around Oph IRS~48.
\cite{AB15} found almost perfectly axisymmetric, ring-like distribution of pebbles around HL~Tau.
The Strategic Explorations of Exoplanets and Disks with Subaru (SEEDS) project \citep{mT09} found a wealth of morphologically diverse structures like spiral arms and rings in an array of protoplanetary disks.
These structures have also been detected with the Very Large Telescope (VLT; see, e.g., \citealt{BJB15}).

To gain more insight into the physical processes at work in these protoplanetary disks, numerical simulations modeling both gas and solid particles are often enlisted given the complexity of such a system.
For example, numerical simulations were used to demonstrate how pebbles/boulders could spontaneously concentrate themselves in a gaseous disk via the streaming instability \citep{JY07,BS10a,YJ14}, to the extent that planetesimal formation is triggered (\citealt{JYM09,JM15}; \citealt{SA15}; Sch\"{a}fer, Yang, \& Johansen, in preparation).
Simulations have also been used to study how dust is trapped within vortices, which could explain the observed lopsided dust concentration in some transition disks \citep{LJ09,ZS14,BZ15,SML16}.
Using simulations of a protoplanetary disk with planet-induced gaps, the axisymmetric dust distribution observed in the HL Tau disk can be reproduced \citep{DP15,DZW15,JL16}.
By matching simulation models with the spiral structures in an observed protoplanetary disk, the mass and orbit of any potentially unseen planet can also be inferred \citep{DZ15,FD15}.

For a solid particle of size less than $\sim$100\,m in a protoplanetary disk, the dominant interaction between the particle and its surrounding gas is via their mutual drag force instead of their mutual gravity \citep[see, e.g.,][]{OMM07,NG10}.
The main tendency of the drag force is to reduce the relative velocity between the gas and the particle exponentially with time, the strength of which can be characterized by the stopping time $t_s$ \citep{fW72,sW77}.
When the collective motion of a swarm of identical solid particles is considered, the time constant of the exponential decay due to this mutual drag force is given by $t_s / (1 + \epsilon)$, where $\epsilon$ is the local solid-to-gas density ratio (see Sections~\ref{SS:asol} and~\ref{SS:unistr}).
Therefore, for any numerical method explicitly integrating a particle-gas system with mutual drag interaction, stability criterion requires that the time step must be less than this time constant at all time.

Depending on the value of this time constant, $t_s / (1 + \epsilon)$, the mutual drag force can become extremely stiff in particular regimes.
Firstly, for particles of sizes less than about the mean free path of the surrounding gas, the stopping time $t_s$ is linearly proportional to the size of the particles \citep{pE24,sW77}.
Hence the smaller the particles, the stiffer the mutual drag force becomes.
As a reference point, $t_s$ is on the order of $10^{-5}$ local Keplerian orbital period for mm-sized particles embedded at 1\,au in the mid-plane of a Minimum Mass Solar Nebula model \citep[e.g.,][]{aY10}.
Secondly, the stronger the local concentration of solid particles, the higher the maximum solid-to-gas density ratio $\epsilon$ and yet again the stiffer the mutual drag force is.
It has been suggested that the mm-sized chondrules ubiquitous in the Solar System were formed in a solid-rich environment, with a density of solids roughly 100 times the background gas density or more \citep{CA06,AG08}.
Moreover, for sedimented layer of particles marginally coupled to the surrounding gas, the maximum local solid-to-gas density ratio in the saturated state of the streaming turbulence can be as high as $\epsilon \sim 10^2$--$10^3$ \citep{JO07,BS10a,YJ14}.
Therefore, either effect or a combination of both can render the time step so short that the computational cost of the simulation model becomes intractable.

A classic approach to treat stiff source terms is to operator split these terms out and use a dedicated method to integrate them with reasonable accuracy and efficiency \citep[e.g.,][]{II08,YK12}.
When considering two-fluid approximation for a particle-gas system with mutual drag interaction, this leads to a system of ordinary differential equations for the drag force without any spatial coupling.
It is then relatively straightforward to integrate this system strictly locally, either with analytical formulas or with numerical methods.
This approach and similar ones have been implemented in Eulerian grid-based schemes \citep{PM06,BT09,fM10,KH13} and in smoothed particle hydrodynamics \citep{LP12,LB14,LB15,BSC15}.

Instead of the two-fluid approximation, however, it is typically preferable to use Lagrangian solid particles to model a particle-gas system \citep{YJ07,BT09,BS10b,fM10,ZS14}.
The exact information of position and velocity carried by each particle better samples the distribution of particles in phase space.
The ability of sampling the velocity distribution of particles is especially important in the study of collisional evolution of solids, both large \citep[e.g.,][]{KI96} and small \citep[e.g.,][]{ZD08}.
Lagrangian particles not only help measure their velocity dispersion, but also can directly be used to predict the collision parameters, neither of which is readily available in the fluid approximation.
Moreover, in the drag-dominated regime, particles do not dynamically equilibrate among themselves due to lack of direct interaction, and thus the effective pressure of particles is virtually zero.
This implies that any spatial variation in the fluid description of particles is connected by a contact discontinuity, which is not trivial to be traced numerically accurately, especially when the contrast is significant.
This issue is absent when using Lagrangian particles.

Despite these advantages of employing Lagrangian solid particles along with Eulerian gas, major difficulties arise for the direct integration of the mutual drag force between gas and particles (see Section~\ref{S:algm}).
The most difficult of these is that the presence of the particles can make the system of equations globally coupled, much like a diffusion equation, with which the temporal solution for any cell depends on the initial conditions for all other cells.
With the diffusion equation, an initial delta function is immediately broadened in time and becomes a Gaussian distribution which is nonzero for the whole domain.
Surprisingly, using the standard particle-mesh approach to compute the mutual drag force in a particle-gas system also induces this property for the connected domain covered by the particles, as we show in Appendix~\ref{S:ns}.
Consider, for example, a one-dimensional grid of gas with uniformly distributed particles.
Suddenly pushing one particle on one side of the domain would drag not only its surrounding gas, but also all of the particles via their intermediate gas cells, up to the particles on the other side of the domain.
Though this effect of global propagation of information diminishes exponentially with distance, it indicates that using the particle-mesh method, the coupling between the gas and the particles is more complex than simple local particle-gas pairs.
While there exist standard numerical techniques to treat a diffusion equation efficiently, e.g., the Crank--Nicolson method and the spectral method, the incongruence between the Lagrangian and the Eulerian descriptions makes these methods not applicable to the particle-gas systems with mutual drag force.
Therefore, the focus has been only on the direct integration of the drag force on the particles without operator splitting the drag force on the gas, as in \cite{BT09}, \cite{BS10b}, and \cite{fM10}.
This approach only relieves the time-step constraint due to small particles, while it remains problematic as strong solid concentration occurs.
Note also that in \cite{BS10b} as well as in \cite{JO07}, an artificial increase of the stopping time $t_s$ is implemented for those cells with high local solid-to-gas density ratios in order to circumvent the time-step constraint, and the numerical accuracy of this approach has not been systematically demonstrated yet.

In this work, we devise a numerical algorithm that effectively disentangles the system of equations for the mutual drag force and allows for its direct integration on a cell-by-cell basis.
For each cell, then, we use an analytical solution to assist in predicting the velocities of the gas and the particles at the next time step, so that the time-step constraint posed by the mutual drag force is lifted.
This algorithm is described in detail in Section~\ref{S:algm}.
To validate our algorithm, we use an extensive suite of benchmarks with known solutions in one, two, and three dimensions, presented in Sections~\ref{S:1d}--\ref{S:3d}, respectively.
Finally, we discuss the generality and possible applications of our algorithm in Section~\ref{S:conc}.

\section{THE ALGORITHM} \label{S:algm}

\begin{figure*}
\begin{center}
\plottwo{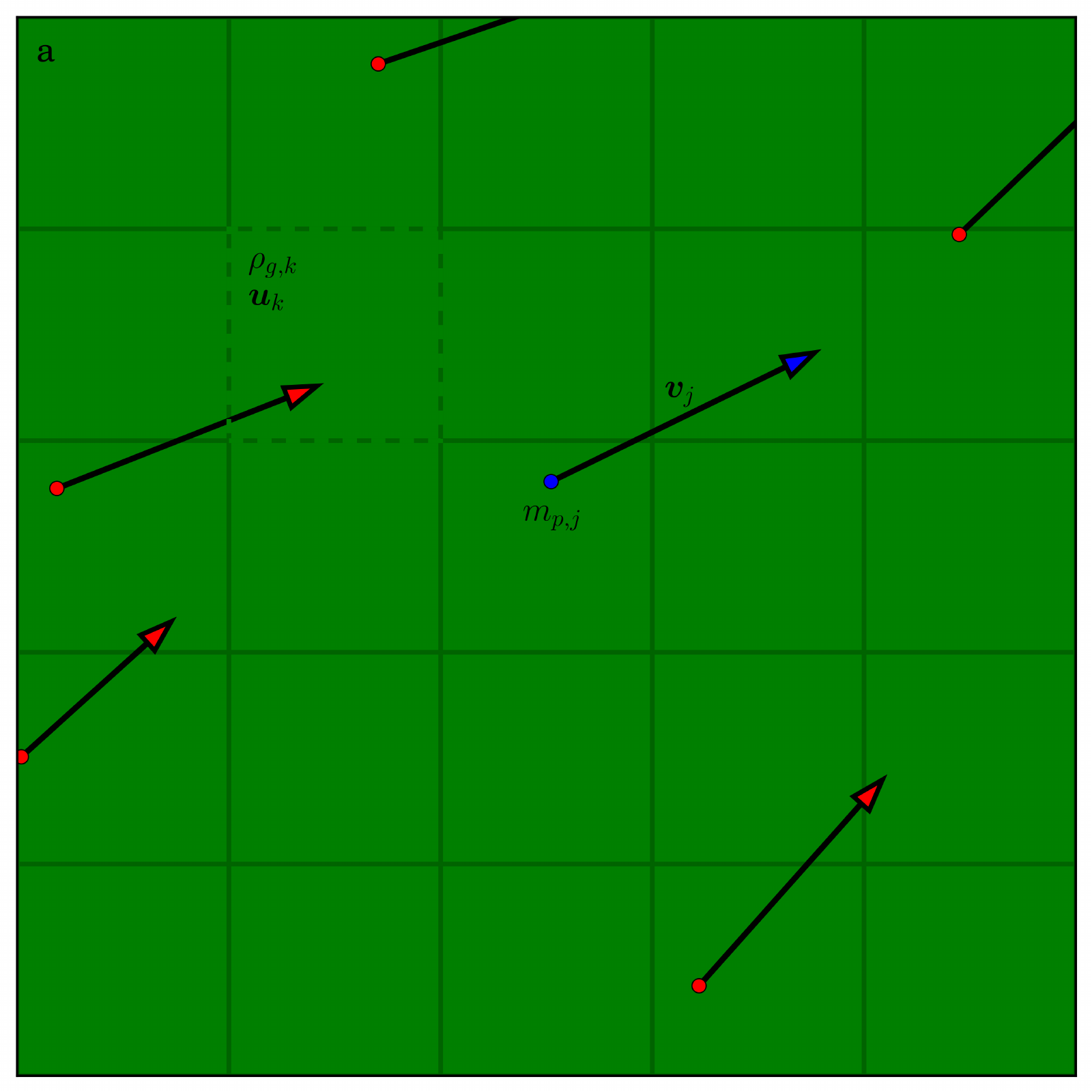}{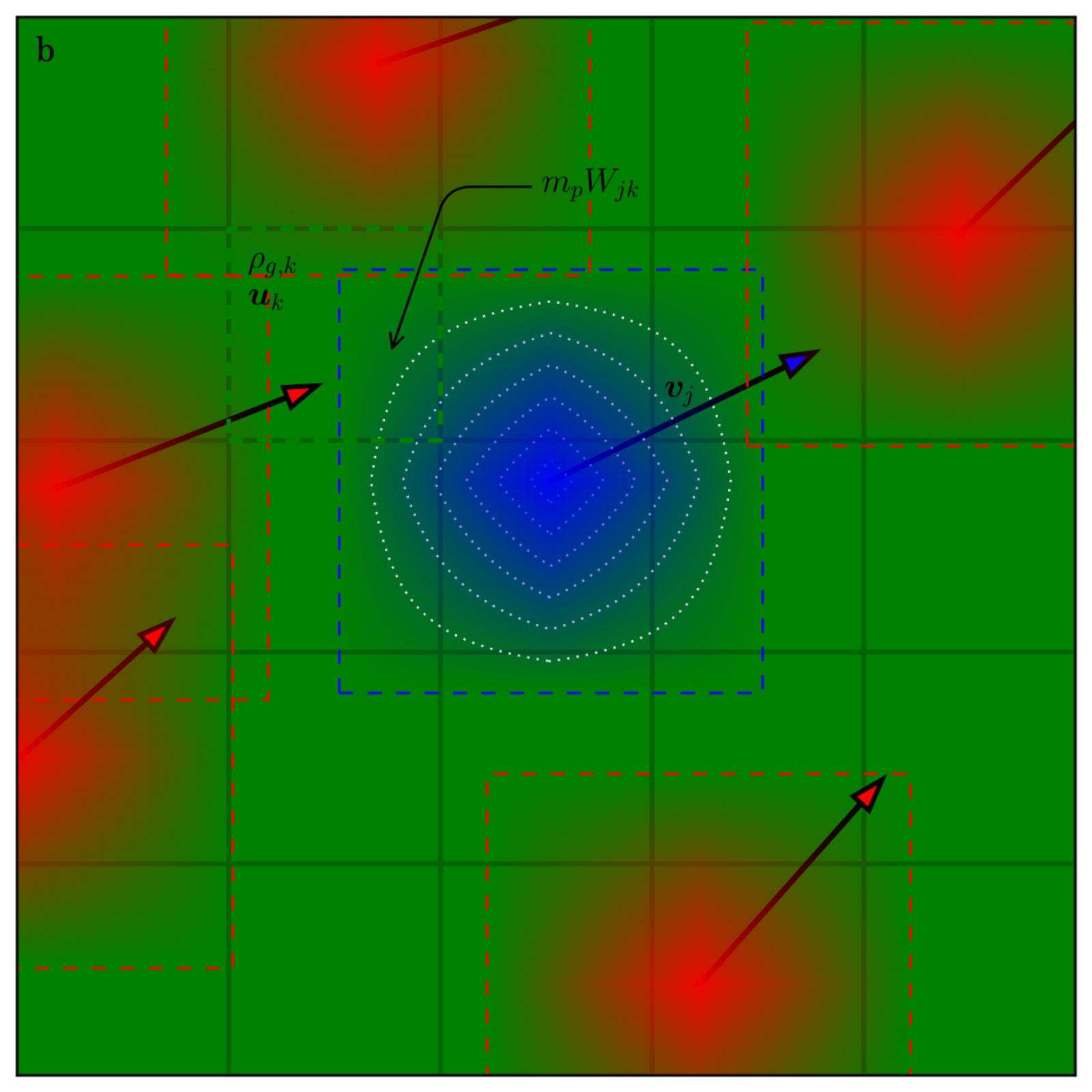}
\caption{Schematic diagrams in two dimensions, demonstrating the particle-mesh scheme we use to integrate the mutual drag force.
(a)~A grid of gas cells with several Lagrangian solid particles.
    The $j$-th particle (shown in blue) has a mass of $m_{p,j}$ and a velocity of $\vec{v}_j$.
    The green dashed line shows the boundary of cell $k$, in which the gas has a density of $\rho_{g,k}$ and a velocity of $\vec{u}_k$.
(b)~Each particle is interpreted as a distributed cloud, with its boundary shown by a blue or red dashed line.
    The white dotted lines are the density contours for the $j$-th particle.
    Each particle cloud is further split by the grid cells into sub-clouds, each of which has the same initial velocity $\vec{v}_j$ as the parent particle.
    The mass of the sub-cloud from the $j$-th particle that is enclosed in cell $k$ is $m_{p,j} W_{jk}$, where $W_{jk} \equiv W(\vec{r}_{p,j} - \vec{r}_k)$.
}
\label{F:pm}
\end{center}
\end{figure*}

To begin with, we consider a system of Eulerian gas and Lagrangian solid particles moving in a differentially rotating disk in a rotating frame.
In addition to the Coriolis force, the centrifugal force, and the external axisymmetric gravitational potential, the gas and each of the particles interact with their mutual drag force.
We further adopt the local-shearing-sheet approximation \citep{GL65}, in which the origin of the frame is located at an arbitrary distance $R$ away from the rotation axis with the $x$-, $y$-, and $z$-axes of the frame constantly in the radial, azimuthal, and vertical directions, respectively, and the frame co-rotates with the disk at the local angular frequency at its origin.
We include also a constant $x$-acceleration $a_x$ to the gas from, e.g., a background pressure gradient, and a vertically varying $z$-acceleration $g_z(z)$ on both the gas and the particles due to, e.g., the vertical component of the central gravity.
Without loss of generality, we assume an isothermal equation of state for the gas with $c_s$ being the speed of sound.
Then the governing equations for the gas read
\begin{align}
&\pder{\rho_g}{t} -
    q\Omega x\pder{\rho_g}{y} +
    \vec{u}\cdot\vec{\nabla}\rho_g +
    \rho_g\vec{\nabla}\cdot\vec{u} = 0,\label{E:gas_cont}\\
&\pder{\vec{u}}{t} -
    q\Omega x\pder{\vec{u}}{y} +
    \vec{u}\cdot\vec{\nabla}\vec{u} =
    a_x\unitvec_x + g_z(z_k)\unitvec_z -
    c_s^2\vec{\nabla}\ln\rho_g\nonumber\\&\qquad -
    2\vec{\Omega}\times\vec{u} + q \Omega u_x \unitvec_y +
    \sum_j\frac{\epsj(\vec{v}_j - \vec{u})}{t_s},\label{E:gas_mom}
\end{align}
while the equations of motion for the particles read
\begin{align}
\tder{\vec{r}_{p,j}}{t} &=
    -q\Omega x_{p,j}\unitvec_y + \vec{v}_j,\label{E:par_vel}\\
\tder{\vec{v}_j}{t} &=
    g_z(z_{p,j})\unitvec_z - 
    2\vec{\Omega}\times\vec{v}_{j} + q \Omega v_{j,x} \unitvec_y +
    \frac{\tilde{\vec{u}}_j - \vec{v}_j}{t_s}.\label{E:par_acc}
\end{align}
In the above equations, $\rho_g$ and $\vec{u}$ are respectively the density and the velocity of the gas at grid point $\vec{r}_k = (x_k, y_k, z_k)$, and $\vec{v}_j$ is the velocity of the $j$-th particle which is located at $\vec{r}_{p,j} = (x_{p,j}, y_{p,j}, z_{p,j})$; see Figure~\ref{F:pm}a.
The constant local angular velocity $\vec{\Omega} = \Omega(R)\unitvec_z$ is parallel to the $z$-axis, and $q = -\mathrm{d}\ln\Omega / \mathrm{d}\ln R$ is the dimensionless shear parameter at radial distance $R$ from the rotation axis, which is $3/2$ for a Keplerian potential.
Both $\vec{u}$ and $\vec{v}_j$ are measured relative to the background shear velocity $-q\Omega x\unitvec_y$ at their respective locations.
The parameter $t_s$ is the stopping time of the mutual drag force between the gas and each of the particles.
For simplicity, and more importantly, for a clean demonstration of our algorithm, we assume $t_s$ is a constant in this work and discuss the possibility of its generalization in Section~\ref{S:conc}.
The remaining variables are $\epsj$ and $\tilde{\vec{u}}_j$: the former is some averaged particle-to-gas density ratio ``perceived'' by the gas in cell $k$ from the $j$-th particle, while the latter is some averaged velocity of the surrounding gas ``perceived'' by the $j$-th particle.

We then operator split out the source terms for the mutual drag force, the rotation/shear, the external radial acceleration for the gas, and the external vertical acceleration for the particles from the full system of equations~\eqref{E:gas_cont}--\eqref{E:par_acc} (see Appendix~\ref{S:os}).
As a result, the two independent systems of equations now read\footnote{The reason that we do not operator split the external vertical acceleration for the gas is because it is more accurate for this term to be in balance with the vertical pressure gradient in the same integrator.  This approach also allows us to maintain the formulation of density stratification subtraction proposed in \citet{YJ14}.}
\begin{subequations} \label{E:hypsys}
\begin{align}
&\pder{\rho_g}{t} -
    q\Omega x\pder{\rho_g}{y} +
    \vec{u}\cdot\vec{\nabla}\rho_g +
    \rho_g\vec{\nabla}\cdot\vec{u} = 0,\label{E:gascon}\\
&\pder{\vec{u}}{t} -
    q\Omega x\pder{\vec{u}}{y} +
    \vec{u}\cdot\vec{\nabla}\vec{u} =
    g_z(z_k)\unitvec_z -
    c_s^2\vec{\nabla}\ln\rho_g,\label{E:gasmom}\\
&\tder{\vec{r}_{p,j}}{t} =
    -q\Omega x_{p,j}\unitvec_y + \vec{v}_j,
\end{align}
\end{subequations}
and
\begin{subequations} \label{E:src}
\begin{align}
\tder{\vec{u}}{t} &=
    a_x\unitvec_x - 
    2\vec{\Omega}\times\vec{u} + q \Omega u_x \unitvec_y +
    \sum_j\frac{\epsj(\vec{v}_j - \vec{u})}{t_s},\label{E:srcgas}\\
\tder{\vec{v}_j}{t} &=
    g_z(z_{p,j})\unitvec_z - 
    2\vec{\Omega}\times\vec{v}_{j} + q \Omega v_{j,x} \unitvec_y +
    \frac{\tilde{\vec{u}}_j - \vec{v}_j}{t_s}.\label{E:srcpar}
\end{align}
\end{subequations}%
The system of equations~\eqref{E:hypsys} consists of the usual Euler equations for fluid dynamics and an ordinary differential equation for the positions of the particles, both of which can be integrated with any standard technique.
Our task is therefore to devise a numerical algorithm to solve the system of equations~\eqref{E:src} without any time-step constraint due to the mutual drag force between the gas and the particles.

Even after the operator splitting, the system of equations~\eqref{E:src} still presents several major difficulties, as mentioned in Section~\ref{S:intro}.
Firstly, the rather loose definitions of $\tilde{\epsilon}_{kj}$ and $\tilde{\vec{u}}_j$ are a manifestation of the distinctly different formalisms of the gas and the solid particles, i.e., Eulerian vs.\ Lagrangian.
Their design is therefore one of the central factors that determine the accuracy as well as the efficiency of the algorithm concerning the mutual drag force.
Secondly, to relieve the time-step constraint posed by both small stopping time $t_s$ and large solid-to-gas density ratio $\epsilon$, the characteristic velocity curves followed by each cell of gas and each particle need to be accurately captured by the numerical method.
This is best achieved by solving the system of equations~\eqref{E:src} simultaneously in the same integrator, given the particle-gas coupling via the mutual drag force.
Worst of all, the density ratio $\tilde{\epsilon}_{kj}$ usually includes all particles within a definite distance to cell $k$, while the velocity $\tilde{\vec{u}}_j$ is usually a weighted average from the surrounding cells around the $j$-th particle.
This implies that all cells by equation~\eqref{E:srcgas} and all particles by equation~\eqref{E:srcpar} are more than likely to be completely coupled via $\tilde{\epsilon}_{kj}$ and $\tilde{\vec{u}}_j$, particularly when the particles are roughly uniformly distributed throughout the computational domain.
These couplings make the solutions for the whole velocity field of the gas and all the particle velocities as a function of time dependent on each other (see also Appendix~\ref{S:ns}).
In order to solve this system efficiently, an implicit numerical method is usually employed, since the method is unconditionally stable against time step.
However, even with an implicit method, the matrix representing the system of equations~\eqref{E:src} is huge, with the total number of cells plus particles on each side, and the inversion of this matrix can be a prohibitive computational task, since it cannot be organized in simple band diagonal form.\footnote{Various works by \cite{BT09}, \cite{BS10b}, and \cite{fM10} did not have this issue, for they only made the solver for particles implicit and thus each particle can be evolved independently without the knowledge of other particles at the same time step.  Nevertheless, as discussed in Section~\ref{S:intro}, this approach does not relieve the time-step constraint due to high local solid-to-gas density ratios.}
In the following subsections, we construct our numerical algorithm with these difficulties in mind.

\subsection{Analytical Solutions to the System of Equations} \label{SS:asol}

In order to proceed, we first conceive the simplest possible scenario to solve the system of equations~\eqref{E:src}, which leads to the nearest-grid-point (NGP) scheme of the particle-mesh method.
In this scheme, each individual cell containing some number of particles is considered independently; the gas in one cell interacts only with the particles inside and vice versa.
As a result, the gas and the particles in a cell do not couple with those in any of the neighboring cells.
This implies that $\epsj = m_{p,j} / \rho_g V$ and $\tilde{\vec{u}}_j = \vec{u}$, where $m_{p,j}$ is the mass of the $j$-th particle and $V$ is the volume of the cell.
With these simplifications, the system of equations~\eqref{E:src} can be solved analytically.

The vertical direction is independent of the other directions, and the solutions for it read
\begin{align}
u_z(t) &= u_z(0) e^{-\pope\tau} + \alpha t +\nonumber\\&\quad
    \left(U_z - \frac{\alpha t_s}{\ope}\right)
    \left[1 - e^{-\pope\tau}\right],\label{E:uzt}\\
v_{j,z}(t) &=
    \left[v_{j,z}(0) + 
          u_z(0)\left(\frac{1 - e^{-\eps\tau}}{\eps}\right)\right]e^{-\tau} +
    \alpha t +\nonumber\\&\quad
    \left(U_z - \frac{\alpha t_s}{\ope}\right)
        \left[1 - e^{-\tau}\left(1 - \frac{1 - e^{-\eps\tau}}
                                          {\eps}\right)\right]\nonumber\\&\quad +
    \left[g_z(z_{p,j}) - \alpha\right]t_s\left(1 - e^{-\tau}\right)
    \label{E:vzt},
\end{align}
where
\begin{equation}
\eps \equiv \sum_j\epsj
\end{equation}
is the total solid-to-gas density ratio in the cell,
\begin{equation}
\tau \equiv \frac{t}{t_s}
\end{equation}
is the number of $e$-folding times for the drag force at $t$,
\begin{equation}
\alpha \equiv \frac{\sum_j \epsj g_z(z_{p,j})}{\ope}\label{E:acm}
\end{equation}
is the vertical center-of-mass acceleration, and
\begin{equation}
U_z \equiv \frac{u_z(0) + \sum_j \epsj v_{j,z}(0)}{\ope}
\end{equation}
is the vertical component of the initial center-of-mass velocity of the particle-gas system.
The first term in equations~\eqref{E:uzt} and~\eqref{E:vzt} is the decaying mode from the initial velocities of the gas and the particles.
The second term represents the center-of-mass motion of the system.
The remaining terms stem from the coupling between the gas and the particles due to the mutual drag force and determine the terminal velocities relative to the center-of-mass motion.

Solving the system of equations~\eqref{E:src} for the horizontal directions is more involved because the Coriolis force and the shear couple the $x$- and $y$-components of the velocities.  Nevertheless, the analytical solutions can still be found and are
\begin{align}
u_x(t) &= \tilde{u}_x +
    \left(U_x\cos\ot + \beta U_y\sin\ot\right)\nonumber\\&\qquad +
    \eps e^{-\pope\tau}\left(V_x\cos\ot + \beta V_y\sin\ot\right),
    \label{E:uxt}\\
u_y(t) &= \tilde{u}_y +
    \left(U_y\cos\ot - \beta^{-1} U_x\sin\ot\right)\nonumber\\&\qquad +
    \eps e^{-\pope\tau}\left(V_y\cos\ot - \beta^{-1}V_x\sin\ot\right),
    \label{E:uyt}\\
v_{j,x}(t) &= \tilde{v}_x +
    \left(U_x\cos\ot + \beta U_y\sin\ot\right)\nonumber\\&\qquad -
    e^{-\pope\tau}\left(V_x\cos\ot + \beta V_y\sin\ot\right)\nonumber\\&\qquad +   
    e^{-\tau}\left(w_{j,x}\cos\ot + \beta w_{j,y}\sin\ot\right),
    \label{E:vxt}\\
v_{j,y}(t) &= \tilde{v}_y +
    \left(U_y\cos\ot - \beta^{-1} U_x\sin\ot\right)\nonumber\\&\qquad -
    e^{-\pope\tau}\left(V_y\cos\ot - \beta^{-1}V_x\sin\ot\right)\nonumber\\&\qquad +
    e^{-\tau}\left(w_{j,y}\cos\ot - \beta^{-1}w_{j,y}\sin\ot\right).
    \label{E:vyt}
\end{align}
The first term in the above equations denotes the equilibrium velocities, which are calculated by
\begin{align}
\tilde{u}_x &\equiv
    \frac{2\eps\tau_s}{\pope^2 + 2(2-q)\tau_s^2}
    \left(\frac{a_x}{2\Omega}\right),\label{E:uxeq}\\
\tilde{u}_y &\equiv
    -\frac{\pope + 2(2-q)\tau_s^2}{\pope^2 + 2(2-q)\tau_s^2}
    \left(\frac{a_x}{2\Omega}\right),\label{E:uyeq}\\
\tilde{v}_x &\equiv
    -\frac{2\tau_s}{\pope^2 + 2(2-q)\tau_s^2}
    \left(\frac{a_x}{2\Omega}\right),\label{E:vxeq}\\
\tilde{v}_y &\equiv
    -\frac{\ope}{\pope^2 + 2(2-q)\tau_s^2}
    \left(\frac{a_x}{2\Omega}\right),\label{E:vyeq}
\end{align}
where $\tau_s \equiv \Omega t_s$ is the dimensionless stopping time, or the Stokes number in the context of a rotating disk.
When the shear parameter $q = 3/2$ and $\Omega = \Omega_K$, the Keplerian frequency, these are simply the Nakagawa--Sekiya--Hayashi (NSH) equilibrium solutions \citep{NSH86}.  The vectors $\vec{U}$ and $\vec{V}$ are defined by
\begin{align}
\vec{U} &\equiv
    \frac{\left[\vec{u}(0) - \tilde{\vec{u}}\right] +
           \sum_j\epsj\left[\vec{v}_j(0) - \tilde{\vec{v}}\right]}{\ope},\\
\vec{V} &\equiv
    \frac{\left[\vec{u}(0) - \tilde{\vec{u}}\right] -
          \eps^{-1}\sum_j\epsj\left[\vec{v}_j(0) - \tilde{\vec{v}}\right]}
         {\ope},
\end{align}
which are respectively the initial center-of-mass velocity of the system and the weighted velocity difference between the gas and the particles, measured with respect to the equilibrium state.
The constant $\beta$ is defined by
\begin{equation}
\beta \equiv \sqrt{\frac{2}{2-q}}
\end{equation}
and
\begin{equation}
\omega \equiv \sqrt{2(2-q)}\Omega
\end{equation}
is the epicycle frequency.
Hence the second term in equations~\eqref{E:uxt}--\eqref{E:vyt} describes the constant, in-phase, epicycle motion of the bulk system, while the third term depicts the decaying, out-of-phase, epicyclic mode of the velocity difference between the gas and the particles.
Finally, the vectors $\vec{w}_j$ are defined by
\begin{equation}
\vec{w}_j \equiv 
    \vec{v}_j(0) - \frac{1}{\eps}
                   \sum_l\tilde{\epsilon}_{kl}
                         \left[\vec{v}_l(0) - \tilde{\vec{v}}\right],
\end{equation}
and thus the last term in equations~\eqref{E:vxt} and~\eqref{E:vyt} denotes the decaying, epicyclic mode of the velocity difference of each individual particle relative to the center-of-mass of the particle system.

It should be clear now what the advantages of operator splitting not only the mutual drag force but also the other source terms, as in equations~\eqref{E:src}, are.
These source terms do not depend on any spatial derivative of the field variables or differences between particles, and an equilibrium state should be preserved numerically when they cooperate in balance.
Hence the equilibrium velocities $\tilde{\vec{u}}$ and $\tilde{\vec{v}}$ inherent in equations~\eqref{E:uxt}--\eqref{E:vyt} are important in guaranteeing that the equilibrium state can be maintained down to the machine precision.\footnote{In practice, the round-off errors can serve as the seeds to ``physical'' instabilities in multi-dimensional models.  However, given that many more $e$-folding times is required for these seeds to grow to appreciable amplitude compared to the dynamical timescale of many models of interest, these errors do not pose a serious issue.}
Note that the hydrostatic equilibrium is on the contrary determined by the system of equations~\eqref{E:gascon}--\eqref{E:gasmom}, or the like, which should be independently maintained by the other integrator given the operator split.
Also with equations~\eqref{E:uxt}--\eqref{E:vyt}, all the epicyclic modes due to the coupling between the gas and the particles are accurately followed in time.
Most importantly, the velocity damping due to mutual drag force can be predicted with a time step of arbitrary size using the solutions, and thus the time-step constraint posed by the drag timescale in either direction can be relieved.

\subsection{Update of the Particle Velocities} \label{SS:uppar}
\begin{figure*}
\begin{center}
\figurenum{1 (cont.)}
\plottwo{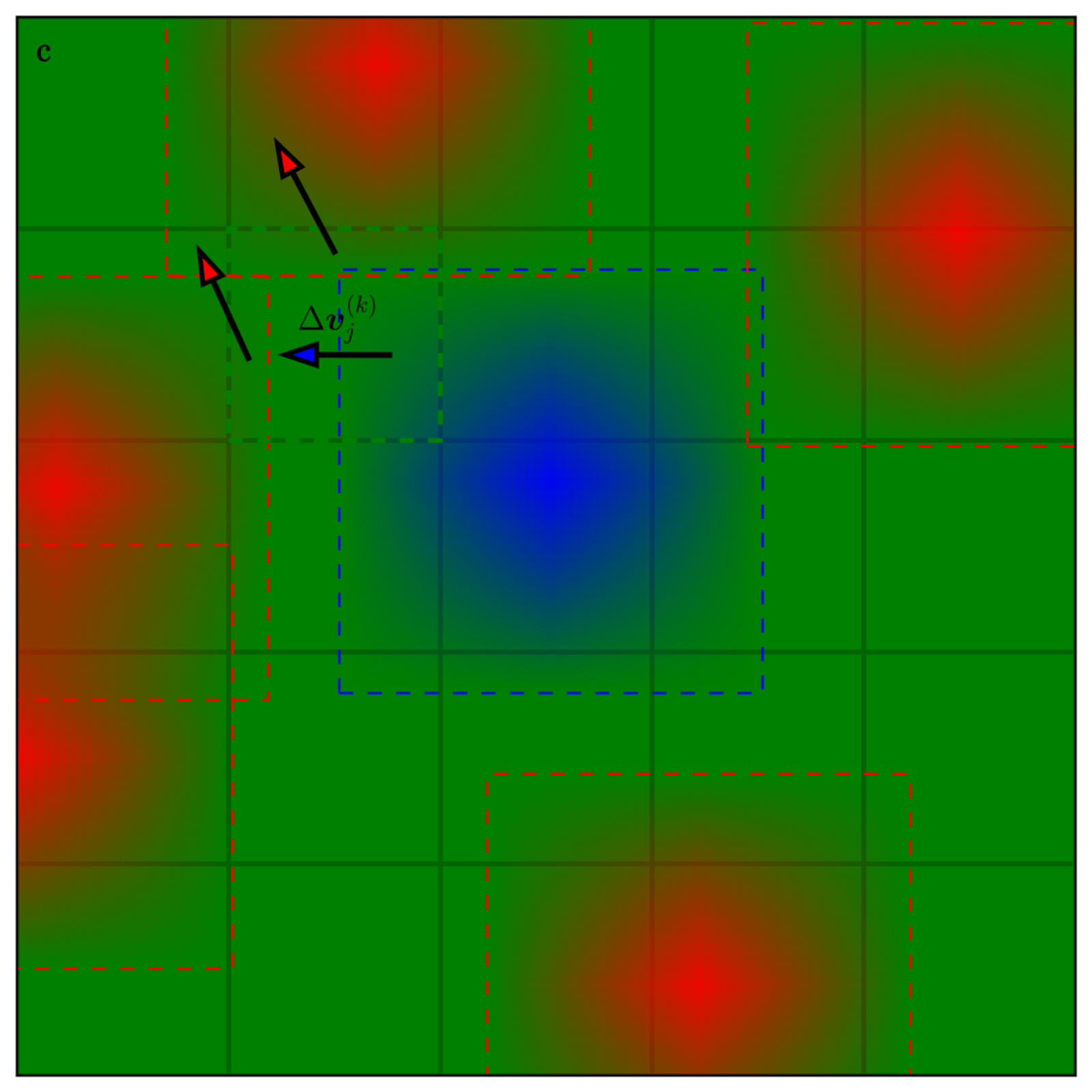}{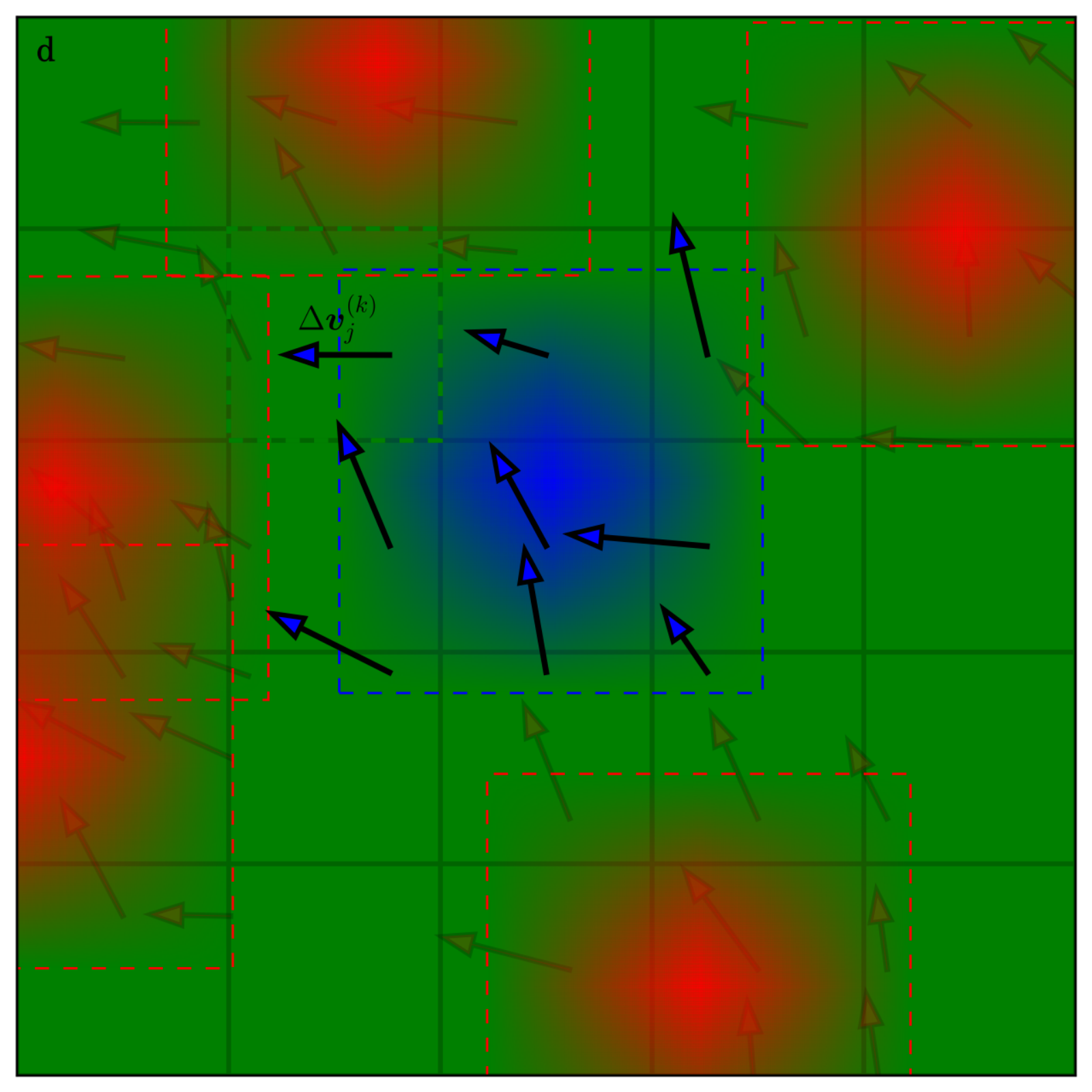}
\plottwo{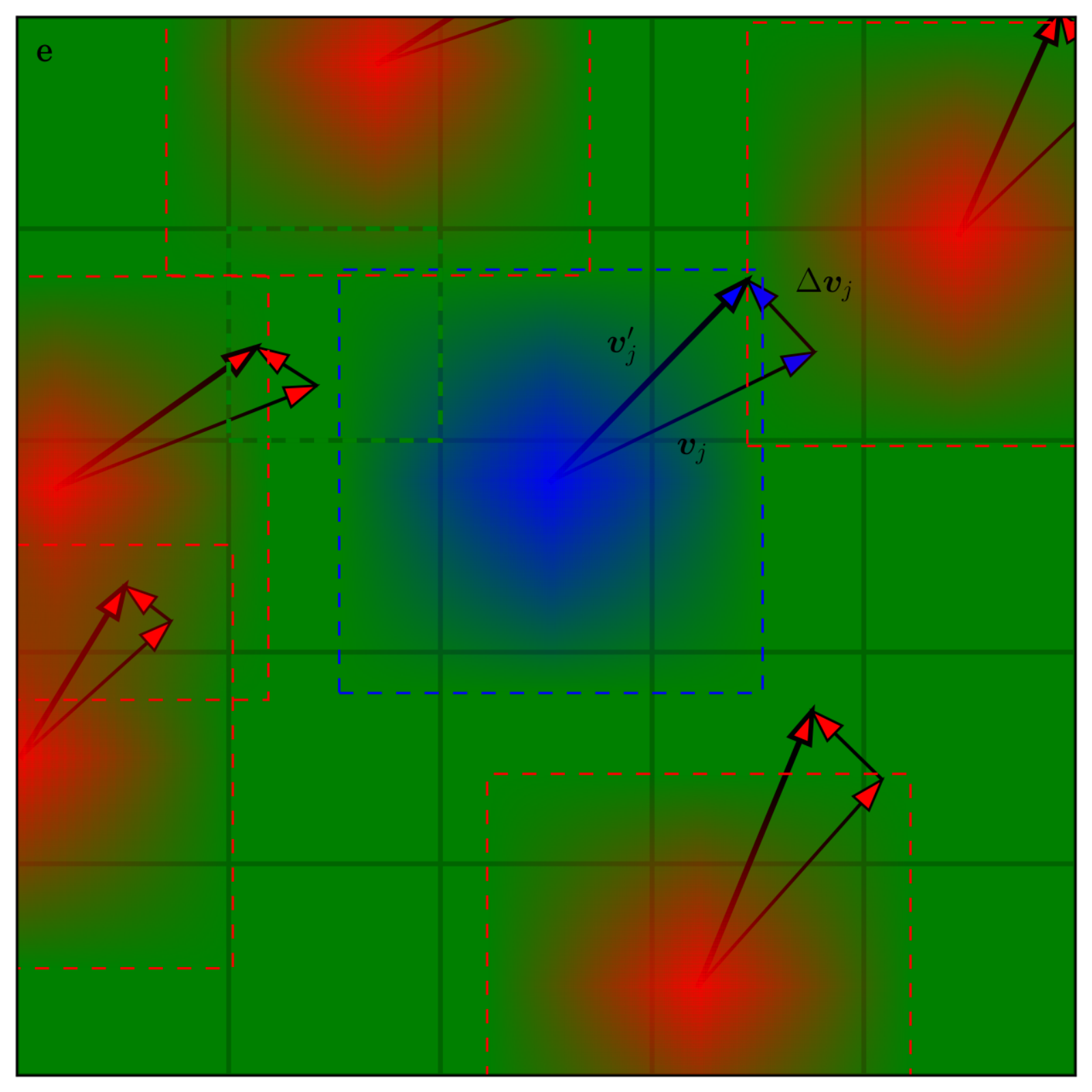}{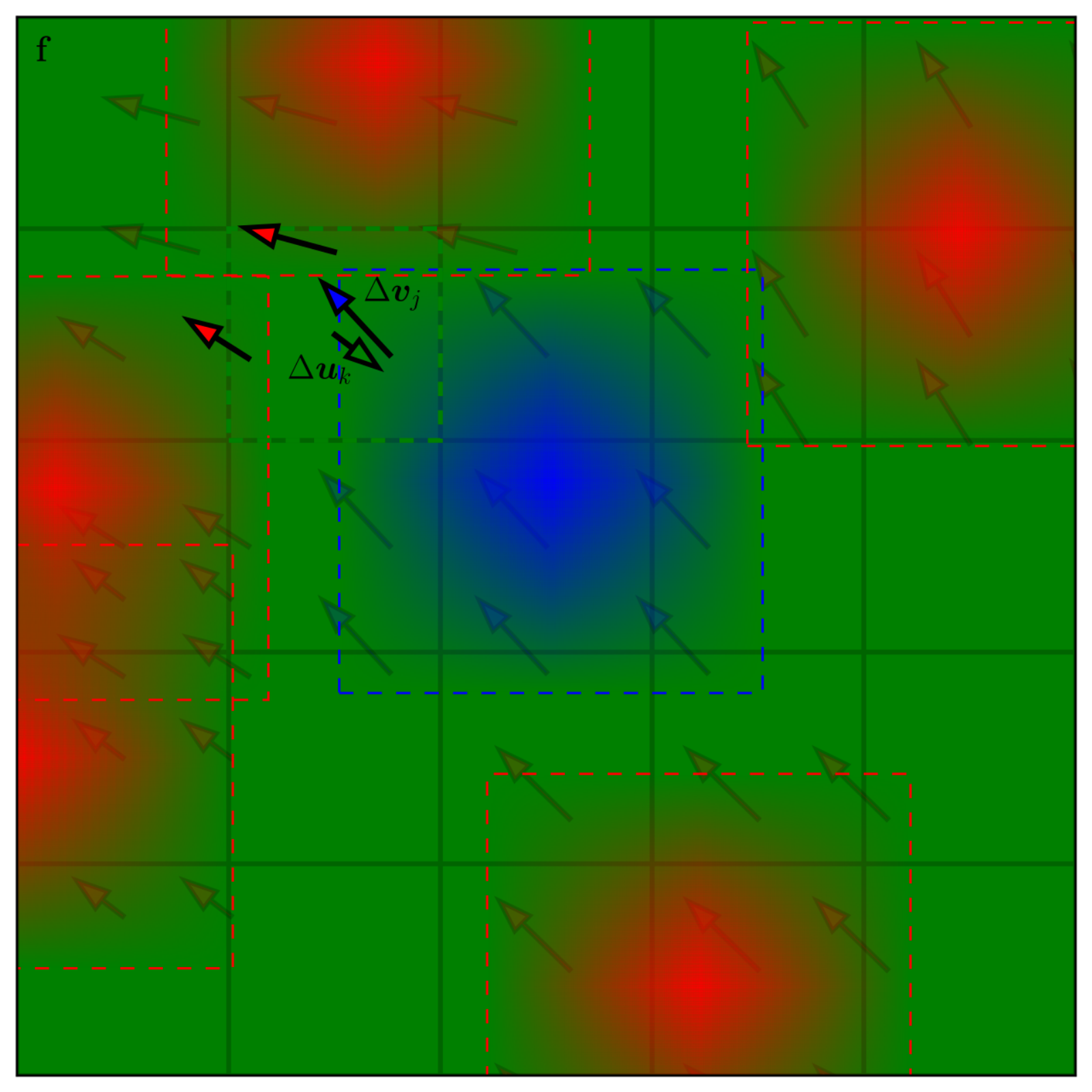}
\caption{
(c)~Within cell $k$, the enclosed sub-clouds are allowed to interact with the gas as well as each other via the mutual drag force, independent of the other sub-clouds in other cells.
    The sub-cloud from the $j$-th particle thus receives a change in velocity $\Delta\vec{v}_j^{(k)}$ over a finite time step $\Delta t$, similarly for the other sub-clouds in the same cell.
(d)~The same operation is applied to every cell, and thus all the sub-clouds undergo a change in velocity.
(e)~The total change in velocity for the $j$-th particle $\Delta\vec{v}_j$ is then the weighted summation of all $\Delta\vec{v}_j^{(k)}$'s of its sub-clouds.
    As a result, the velocity of the $j$-th particle becomes $\vec{v}'_j$ at the next time step, similarly for all other particles.
(f)~To find the change in the velocity field of the gas, we apply once again the interpretation of the sub-clouds, except that each sub-cloud from the $j$-th particle now has the same change of velocity $\Delta\vec{v}_j$ as its parent particle cloud.
    Since the change in total momentum for cell $k$ is known, the change of gas velocity $\Delta\vec{u}_k$ in the cell can be computed from the momentum changes of the sub-particle clouds in the cell.
    We dub this last procedure ``particle-mesh back reaction''.}
\end{center}
\end{figure*}

Even though the NGP scheme is simple and easy to implement, it is not suitable in many circumstances and  a higher-order particle-mesh scheme is usually desirable.
With the NGP scheme, each particle only interacts with the gas at the nearest grid point, so the particle behaves as if sitting at the grid point and the significance of its positional information within the cell is lost.
Moreover, the mass and the momentum density fields sampled via the particles are prone to be overwhelmed by the Poisson noise.
To have a 1\% sensitivity, at least $10^4$ particles are required in each cell.
On the other hand, any higher-order interpolation scheme that utilizes the positional information of the particles drastically overturns the situation; as less as one particle per cell is sufficient to describe a signal of arbitrarily low amplitude \citep{YJ07}.
Therefore, it is necessary to incorporate any particle-mesh scheme of choice into the solutions discussed in Section~\ref{SS:asol}.

To achieve this, the solid-to-gas density ratio contributed by the $j$-th particle in the cell at $\vec{r}_k$ should be generalized to
\begin{equation}
\epsj = \frac{m_{p,j}W(\vec{r}_k - \vec{r}_{p,j})}{\rho_g V}\label{E:epsj},
\end{equation}
where $W(\vec{r})$ is the weight function of the particle-mesh scheme in question.
More often than not, the weight function $W(\vec{r})$ has a physical interpretation \citep[e.g.,][]{HE88}; see Figure~\ref{F:pm}b.
For example, in the cloud-in-cell (CIC) scheme, the mass of each particle is distributed uniformly in a rectangular box of the cell size that is centered at the particle, i.e., a uniform rectangular ``particle cloud''.
In the triangular-shaped-cloud (TSC) scheme, each dimension of the particle cloud is doubled and the mass of the particle is distributed non-uniformly with a density peak at the cloud center and zero density at the cloud boundary.
This scheme is so named since the density profile of the particle cloud resembles an isosceles triangle when viewed at the cloud center along any of the coordinate directions.
The density function of the NGP scheme in this interpretation is simply a delta function.
Hence the term $m_{p,j}W(\vec{r}_k - \vec{r}_{p,j})$ in equation~\eqref{E:epsj} can be interpreted as the fraction of the mass of the $j$-th ``particle cloud'' that is enclosed by the cell at $\vec{r}_k$ and is treated as part of the ``particle fluid'' in the cell.

Guided by this interpretation of equation~\eqref{E:epsj}, we have the following proposition to update the velocities of the particles.
\begin{enumerate}
\item Split each particle cloud into multiple sub-clouds and distribute them into the surrounding cells according to equation~\eqref{E:epsj}.
Each sub-cloud has the same initial velocity as their parent particle.
We denote the velocity of the sub-cloud at cell $k$ as $\vec{v}_j^{(k)}$ and thus $\vec{v}_j^{(k)}(0) = \vec{v}_j(0)$ for all $k$.
See Figure~\ref{F:pm}b.
\item For each cell $k$, treat the gas and all the sub-cloud inside as a multi-fluid system.
Hence identify $\vec{v}_j$ as $\vec{v}_j^{(k)}$ and $\tilde{\vec{u}}_j$ as $\vec{u}$ in equations~\eqref{E:src}.
\item Find the velocity changes $\Delta\vec{v}_j^{(k)} \equiv \vec{v}_j^{(k)}(t) - \vec{v}_j^{(k)}(0)$ at time $t$ from the analytical solutions in equations~\eqref{E:vzt}, \eqref{E:vxt}, and~\eqref{E:vyt}.
See Figures~\ref{F:pm}c and~d.\label{dvjk}
\item After all cells are integrated, collect the momentum changes of the sub-clouds back to the parent particles:
\begin{equation} \label{E:dvj}
\Delta\vec{v}_j
\equiv \vec{v}_j(t) - \vec{v}_j(0)
= \sum_k W(\vec{r}_k - \vec{r}_{p,j})\Delta\vec{v}_j^{(k)}.
\end{equation}
\item Update the velocities of the particles by $\vec{v}_j(t) = \vec{v}_j(0) + \Delta\vec{v}_j$.
See Figure~\ref{F:pm}e.
\end{enumerate}
Essentially, this procedure decouples the coupled system of equations~\eqref{E:src} and makes it possible to conduct the integrations on a cell-by-cell basis, and hence substantially reduces the amount of computational work required, as discussed in the introduction of this section.

We note that the procedure outlined above is consistent with standard particle-mesh interpolation.
Since the momentum change of an individual particle is the sum of the momentum changes of its sub-clouds,
\begin{align}
\tder{\vec{v}_j}{t}
&= \sum_k W(\vec{r}_k - \vec{r}_{p,j})\tder{\vec{v}_j^{(k)}}{t}\nonumber\\
&= \sum_k W(\vec{r}_k - \vec{r}_{p,j})\times\nonumber\\&\
    \left[g_z(z_{p,j})\unitvec_z - 
    2\vec{\Omega}\times\vec{v}_j^{(k)} + q \Omega v_{j,x}^{(k)} \unitvec_y +
    \frac{\vec{u}_k - \vec{v}_j^{(k)}}{t_s}\right]\nonumber\\
&= g_z(z_{p,j})\unitvec_z - 
    2\vec{\Omega}\times\vec{v}_j + q \Omega v_{j,x} \unitvec_y +\nonumber\\&\qquad
    \frac{\sum_k W(\vec{r}_k - \vec{r}_{p,j})\vec{u}_k - \vec{v}_j}{t_s},
    \label{E:pmgas}
\end{align}
where $\vec{u}_k$ is the gas velocity at $\vec{r}_k$.
By comparing this with equation~\eqref{E:srcpar}, it can be seen that $\tilde{\vec{u}} = \sum_k W(\vec{r}_k - \vec{r}_{p,j})\vec{u}_k$, which proves that the gas velocity experienced by the particle is the standard particle-mesh interpolation from its surrounding cells.

In principle, the gas velocity in cell $k$ at time $t$ can be similarly obtained along with Step~\ref{dvjk} above via the analytical solutions in equations~\eqref{E:uzt}, \eqref{E:uxt}, and~\eqref{E:uyt}, and thus can be updated directly.
As will be shown in Section~\ref{SS:lsi}, the growth rate of a linear mode for the streaming instability indeed converges with resolution using this approach.
However, the convergence rate is relatively poor compared with that using the explicit integration.
In the next subsection, we devise further steps for the update of the gas velocity that significantly improve the benchmarks.

\subsection{Update of the Gas Velocity} \label{SS:upgas}

The reason for the relatively poor performance of directly using the analytical solutions to update the gas velocity after the steps described in Section~\ref{SS:uppar} is that this approach remains local from the perspective of the gas.
Even though the gas in a cell receives sub-particle clouds from the surrounding cells, it does not interact with the neighboring gas.
This can be seen in the system of equations~\eqref{E:src} with the substitutions $\vec{v}_j \rightarrow \vec{v}_j^{(k)}$ and $\tilde{\vec{u}} \rightarrow \vec{u}$, where the gas velocity $\vec{u}$ in cell $k$ is the sole state variable for the gas and no coupling for gas velocities between different cells exists.
In reality, however, the gas in neighboring cells should couple via the drag force with the interpenetrating particle clouds as interpreted in the particle-mesh method.

This missing coupling can be remedied, as inspired by the algorithm suggested in \cite{YJ07} for distributing the back reaction of the drag forces from particles to gas.
We note that the velocity change of each particle $\Delta\vec{v}_j$ acquired by the steps in Section~\ref{SS:uppar} contains all the information of the mutual drag force between the particle and the surrounding gas.
That is, the particle has sampled the spatial variation in the velocity field of the gas to determine its own velocity change, as shown in equation~\eqref{E:pmgas}.
This process can then be reversed since the mutual drag force forms an action-reaction pair.
Each particle can be considered now as a unified particle cloud undergoing a momentum change $m_{p,j}\Delta\vec{v}_j$ instead of a group of independent sub-clouds.
This momentum change can then be redistributed onto the grid by standard particle-mesh assignment.
See Figure~\ref{F:pm}f.

One more difficulty remains, though, since additional source terms are included in our system and thus the total momentum of the gas and particles in each cell is not conserved.
This difficulty can be resolved with the center-of-mass frame approach, in which the mutual drag force cancels out.
Combining equations~\eqref{E:srcgas}, \eqref{E:srcpar}, and~\eqref{E:acm} gives
\begin{equation}
\tder{\ucm}{t} =
    \frac{a_x}{\ope}\unitvec_x + \alpha\unitvec_z -
    2\vec{\Omega}\times\ucm + q\Omega\bar{u}_x\unitvec_y,\label{E:com}
\end{equation}
where
\begin{equation}
\ucm(t) \equiv \frac{\vec{u}(t) + \sum_j\epsj\vec{v}_j(t)}{\ope}
\end{equation}
is the center-of-mass velocity.
Equation~\eqref{E:com} can be analytically integrated and gives the change at time $t$ as
\begin{equation}
\Delta\ucm = \frac{\Delta\vec{u} + \sum_j\epsj\Delta\vec{v}_j}{\ope}
\label{E:ducm}
\end{equation}
with
\begin{align}
\Delta\bar{u}_x &= -U_x(1 - \cos\ot) + \beta U_y\sin\ot,\\
\Delta\bar{u}_y &= -U_y(1 - \cos\ot) - \beta^{-1}U_x\sin\ot,\\
\Delta\bar{u}_z &= \alpha t.
\end{align}
Since $\Delta\vec{v}_j$ are known, equation~\eqref{E:ducm} can be rearranged to find the velocity change of the gas $\Delta\vec{u}$ (Figure~\ref{F:pm}f).
This completes our algorithm.

\subsection{Boundary Conditions and Implementation}

As a final note, the sheared periodic boundary conditions for the local-shearing-sheet approximation \citep{BN95,HGB95} require some attention in our algorithm.
For the Eulerian description, they state that $f(x,y,z) = f(x + L_x, y - q\Omega L_x t, z)$, where $f$ is any of the dynamical fields, $L_x$ is the radial dimension of the computational domain, and $t$ is the time at the beginning of each time step instead of the size of a time step used liberally in the previous subsections.
Note, however, that our algorithm completely decouples the gas fields and no simultaneous information in any pair of adjacent cells is needed in any of our steps.
On the other hand, all of the coupling is achieved via the splitting of the particle ``clouds''.
The only place that the boundary conditions (as well as domain decomposition in parallel computing) are required, then, is in the particle-mesh weight function $W(\vec{r})$, specifically in equations~\eqref{E:epsj}, \eqref{E:dvj}, and~\eqref{E:ducm}.
Therefore, the radial boundary conditions can simply be executed by shifting the positions of the particles near the radial boundaries by
\begin{subequations} \label{E:parbc}
\begin{align}
x'_{p,j} &= x_{p,j} \pm L_x,\\
y'_{p,j} &= y_{p,j} \mp q\Omega L_x t,\label{E:parbcy}
\end{align}
\end{subequations}%
when these particles are cast into the other side of the boundary, in which the upper/lower sign is taken for the left/right boundary.\footnote{Special care needs to be taken in equation~\eqref{E:parbcy} to fold the $y$-coordinate into the limits of the ghost cell a particle is sent to, by applying the azimuthal periodicity $f(x,y,z) = f(x, y + L_y, z)$, where $L_y$ is the azimuthal size of the computational domain.}
The azimuthal and the vertical boundary conditions for our algorithm can be similarly implemented by revising equations~\eqref{E:parbc}.
All the other properties of the particles remain unchanged.
We note that this approach also eliminates the need of interpolation as required in the implementation of \cite{YJ07}, the latter of which introduces additional numerical errors near the radial boundaries in particle-mesh assignment.

We have implemented our algorithm in the \textsc{Pencil Code}, a high-order finite-difference simulation code for astrophysical fluids and particles \citep{BD02}.\footnote{The \textsc{Pencil Code} is publicly available at the following website: \texttt{\url{http://pencil-code.nordita.org/}}.}
The code employs sixth-order centered differences in space and third-order Runge--Kutta integration in time.
The system of equations~\eqref{E:src} is operator split out of the Runge--Kutta steps and thus separately integrated by the algorithm described in this section.
Throughout this work, we restrict ourselves to the TSC weight function, with which the interpolation error is of second order in cell size \citep{YJ07}.
In the following, we validate the algorithm as well as our implementation on several systems with known solutions.

\section{ONE-DIMENSIONAL TESTS} \label{S:1d}
\subsection{Sedimentation --- Damped Harmonic Oscillation} \label{SS:osc}

Sedimentation of solid particles towards the mid-plane of a gas disk is one of the most important topics in the theory of planet formation.
It is considered to be the first process in the core accretion scenario, creating a dense layer of solid materials that can later concentrate and become seeds of planetary cores \citep{vS69,GW73}.
The degree of sedimentation intimately couples with the gas dynamics, especially in turbulent disks, and thus numerical simulations are often required \citep{CFP06,JHK06,TB10}.
We hereby use a simple form of the sedimentation process as the first benchmark against our integrator.

We consider a single particle moving vertically through a stationary gas.
The particle undergoes a gravitational acceleration of the form $g_z(z_p) = -\omega_0^2 z_p$, where $\omega_0$ is the vertical natural frequency, and the gas drag of stopping time $t_s$.
The equation of motion for the particle is then
\begin{equation} \label{E:dho}
\tdder{z_p}{t} + \frac{1}{t_s}\tder{z_p}{t} + \omega_0^2 z_p = 0.
\end{equation}
This system is the well-known damped harmonic oscillator, and its analytical solutions are readily available.
Using our algorithm, equation~\eqref{E:dho} is equivalently being operator split into two separate equations as
\begin{align}
\tder{z_p}{t} &= v_z,\label{E:dho1}\\
\tder{v_z}{t} &= -\omega_0^2 z_p - \frac{v_z}{t_s}\label{E:dho2},
\end{align}
the first of which is in the Runge--Kutta integrator\footnote{The solution to equation~\eqref{E:dho1} is simply $z_p(t_0 + t) = z_p(t_0) + v_z(t_0) t$ and thus any integrator of order at least one renders this solution.} while the second is in the split integrator of Section~\ref{S:algm}.
Note that in this case the mass of the particle $m_p$ is effectively zero so that the gas is unaffected and remains stationary.
The particle is released at a height of $z_{p,0}$ at rest, i.e., $z_p(0) = z_{p,0}$ and $v_z(0) = 0$.
We integrate this system with both the Godunov and the Strang splitting methods, which are formally first- and second-order accurate, respectively (see Appendix~\ref{S:os}).

Figure~\ref{F:oscex} compares the numerical and the analytical solutions for the cases of a simple harmonic ($t_s \rightarrow \infty$), underdamped ($\omega_0 t_s > 1/2$), critically damped ($\omega_0 t_s = 1/2$), and over-damped ($\omega_0 t_s < 1/2$) oscillator.
We use a fixed and unusually large time step of $\Delta t = \omega_0^{-1}$ to highlight the numerical errors in the comparison.
Even with such a large time step, the numerical solutions agree reasonably well with the analytical ones, especially for more highly damped systems.
The Strang splitting does perform better than the Godunov splitting in particle position, but no appreciable difference appears in particle velocity between the two splitting methods.
In any case, dispersive errors do exist in both position and velocity for the case of oscillatory systems.
However, no diffusive errors exist in either variable, and having this property is important in accurately establishing the scale height of the particle layer, which is one of the critical factors in driving planet formation.
Finally, notice that although the time step is much larger than the stopping time $t_s$ in the over-damped system and thus the initial acceleration of the particle is not resolved, the numerical solution still accurately captures the terminal speed at the very first time step, and even more so at later times.
\begin{figure*}
\begin{center}
\plotone{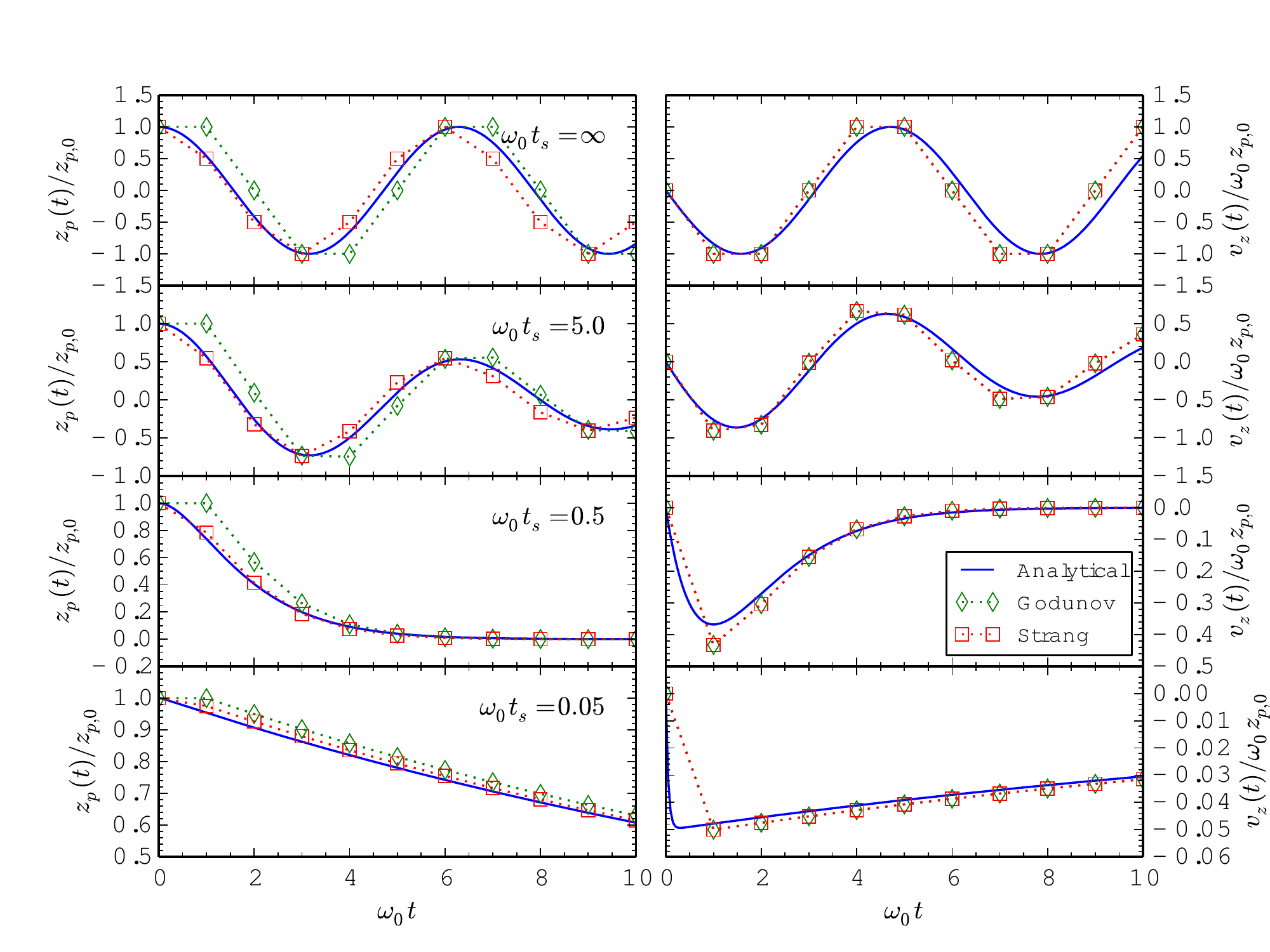}
\caption{Comparison between analytical and numerically obtained solutions for a damped harmonic oscillator.
The left column shows the position of the particle, while the right column shows its velocity, both of which are as a function of time.
The rows, from top to bottom, demonstrate the cases of an undamped ($\omega_0 t_s = \infty$), underdamped ($\omega_0 t_s = 5$), critically damped ($\omega_0 t_s = 0.5$), and over-damped ($\omega_0 t_s = 0.05$) system, respectively.
The solid lines are the analytical solutions, while the diamond and the square symbols denote the numerical solutions using the Godunov and the Strang splitting methods, respectively.
Note that an unusually large time step of $\Delta t = \omega_0^{-1}$ is used to highlight the numerical errors.}
\label{F:oscex}
\end{center}
\end{figure*}

\begin{figure*}
\begin{center}
\plottwo{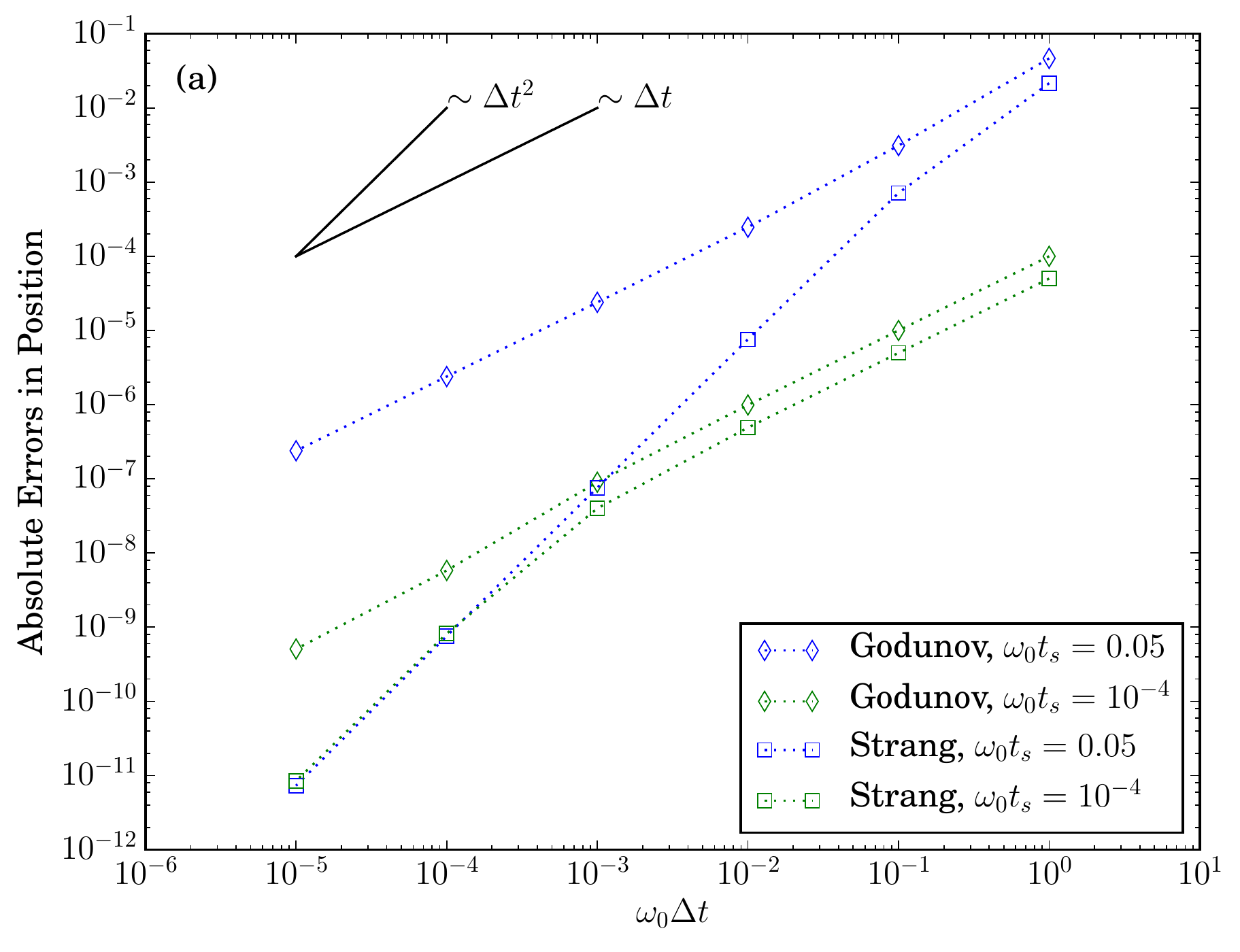}{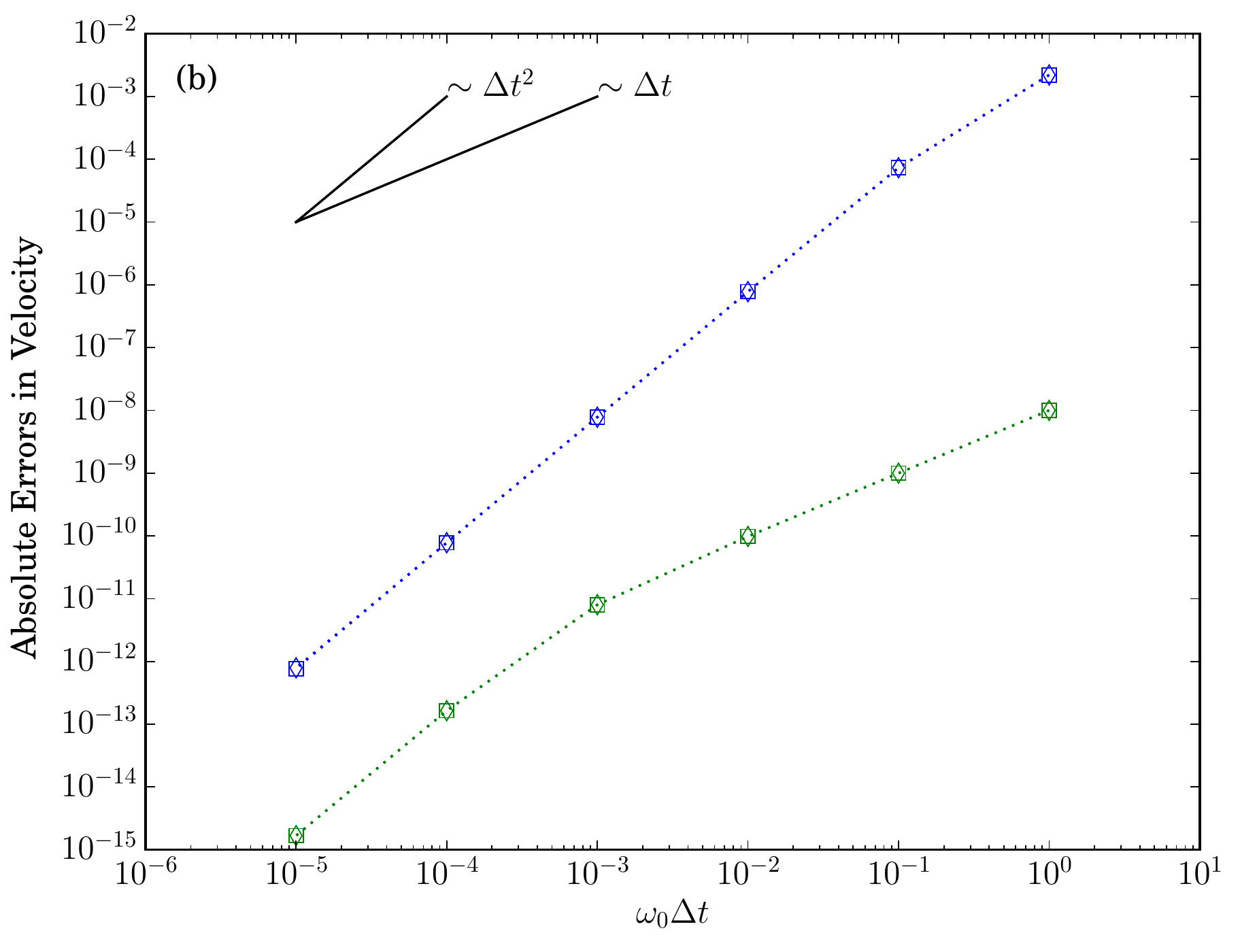}
\caption{Absolute errors in (a)~position and (b)~velocity as a function of time step $\Delta t$ for two over-damped harmonic oscillators.
The position is normalized by the initial position $z_{p,0}$ while the velocity is normalized by $\omega_0 z_{p,0}$, where $\omega_0$ is the natural frequency of the oscillator.
The errors are measured at a fixed final time of $t = \omega_0^{-1}$.
The diamond and the square symbols are the results using the Godunov and the Strang splitting methods, respectively.
The blue and the green colors denote a stopping time of $t_s = 0.05\omega_0^{-1}$ and $t_s = 10^{-4}\omega_0^{-1}$, respectively.
The solid lines guide the slopes for first- or second-order convergence.}
\label{F:oscerr}
\end{center}
\end{figure*}

Figure~\ref{F:oscerr} demonstrates the accuracy and convergence properties of our algorithm for two over-damped harmonic oscillators, one with $\omega_0 t_s = 0.05$ and the other with $\omega_0 t_s = 10^{-4}$.
We cover a wide range of time step $\Delta t$ so that both $\Delta t \gg t_s$ and $\Delta t \ll t_s$ regimes are included.
For the errors in position, the Godunov splitting shows the expected first-order convergence.
On the other hand, the Strang splitting shows the expected second-order convergence only for small $\Delta t$ while first-order convergence for large $\Delta t$, the latter of which might be due to the unresolved initial acceleration of the particle.
The transition occurs at $\Delta t \simeq t_s$, and it seems that the error approaches the same asymptote towards small $\Delta t$ irrespective of the stopping time.
Nevertheless, the Strang splitting is indeed more accurate in position than the Godunov splitting, albeit only slightly at large time step.
For the errors in velocity, there exists no difference between the Godunov and the Strang splittings, which is consistent with Figure~\ref{F:oscex}, and this property does not depend on the stopping time.
Similar to the convergence in position with the Strang splitting, the convergence in velocity for both splittings shows first order for large $\Delta t$ and second order for small $\Delta t$, and the transition occurs at $\Delta t \simeq t_s$.
In any case, the smaller the stopping time, the more accurate the results are at any given time step.
Note that it is never stable to integrate this system explicitly with $\Delta t \gtrsim t_s$, which is the regime of interest in this work.
Furthermore, in a typical model, one usually operates on the range $10^{-3} \lesssim \omega_0 \Delta t \lesssim 10^{-1}$.
Hence we consider these results to be fairly accurate.

\subsection{Uniform Streaming} \label{SS:unistr}

We next consider the interpenetrating streaming motions between uniform gas and uniformly distributed particles.
It is the same test performed by \cite{BS10b}, in which it was called particle-gas deceleration test, and it is also the linear-drag case of the \textsc{dustybox} suite presented by \cite{LP11}.
In this scenario, the system of equations reads
\begin{align}
\tder{u}{t} &= \epsilon\frac{v - u}{t_s},\\
\tder{v}{t} &= \frac{u - v}{t_s},
\end{align}
where $u$ and $v$ are the velocities of the gas and the particles, respectively, and $\epsilon$ is the constant solid-to-gas density ratio.
With the initial conditions $u(0) = u_0$ and $v(0) = v_0$, the solutions for the velocities are
\begin{align}
u(t) &= u_0 e^{-(1+\epsilon)t/t_s} +
         U_0\left[1 - e^{-(1+\epsilon)t/t_s}\right],\label{E:usu}\\
v(t) &= v_0 e^{-(1+\epsilon)t/t_s} +
         U_0\left[1 - e^{-(1+\epsilon)t/t_s}\right],\label{E:usv}
\end{align}
where $U_0 \equiv (u_0 + \epsilon v_0) / (1 + \epsilon)$ is the center-of-mass velocity.
The displacement for each of the particles is given by
\begin{align}
S(t) &\equiv x_p(t) - x_p(0)\nonumber\\
     &= \frac{(v_0 - U_0)t_s}{1 + \epsilon}
        \left[1 - e^{-(1+\epsilon)t/t_s}\right] + U_0 t.\label{E:uss}
\end{align}
In the center-of-mass frame, $U_0 = 0$.
The only relevant scales of time and velocity in this system are the stopping time $t_s$ and the speed of sound $c_s$, respectively.
Hence the time and all the velocities can be normalized by these two scales, and this in turn fixes the length scale at $c_s t_s$.

To test this system with our algorithm, we set up a one-dimensional, periodic grid of gas and uniformly distributed Lagrangian particles.
The computational domain has a length of $100c_s t_s$ so that a time step greater than the stopping time $t_s$ can be covered in the test.
We allocate one particle at the center of each cell.
The particles have an initial velocity of $v_0 = +c_s$ while the gas has a uniform initial velocity of $u_0 = -c_s$, i.e., the gas and the particles move in opposite directions.
The errors are measured at a fixed final time of $t_f = 2t_s$, and the cell size $\Delta x$ is varied to test the numerical convergence.
We note that for $\Delta x = 10c_s t_s$, it takes exactly one time step to reach $t = t_f$.
Finally, we experiment with a wide range of solid-to-gas density ratio $\epsilon$, which is the only free parameter in the system.

We find that the final velocities in all cases are accurate to the analytical solutions, equations~\eqref{E:usu} and~\eqref{E:usv}, close to the machine precision.
Although the velocities do not exactly remain uniform, the noise introduced is close to the machine-precision level.
Only slight increase in the noise level can be observed for high resolutions and thus small time steps, and this can be attributed to the round-off errors.
The high degree of accuracy in velocities is hence not surprisingly due to our analytical integration of the velocities in Section~\ref{SS:asol}, assisted by the high degree of uniform motion.

\begin{figure}
\begin{center}
\plotone{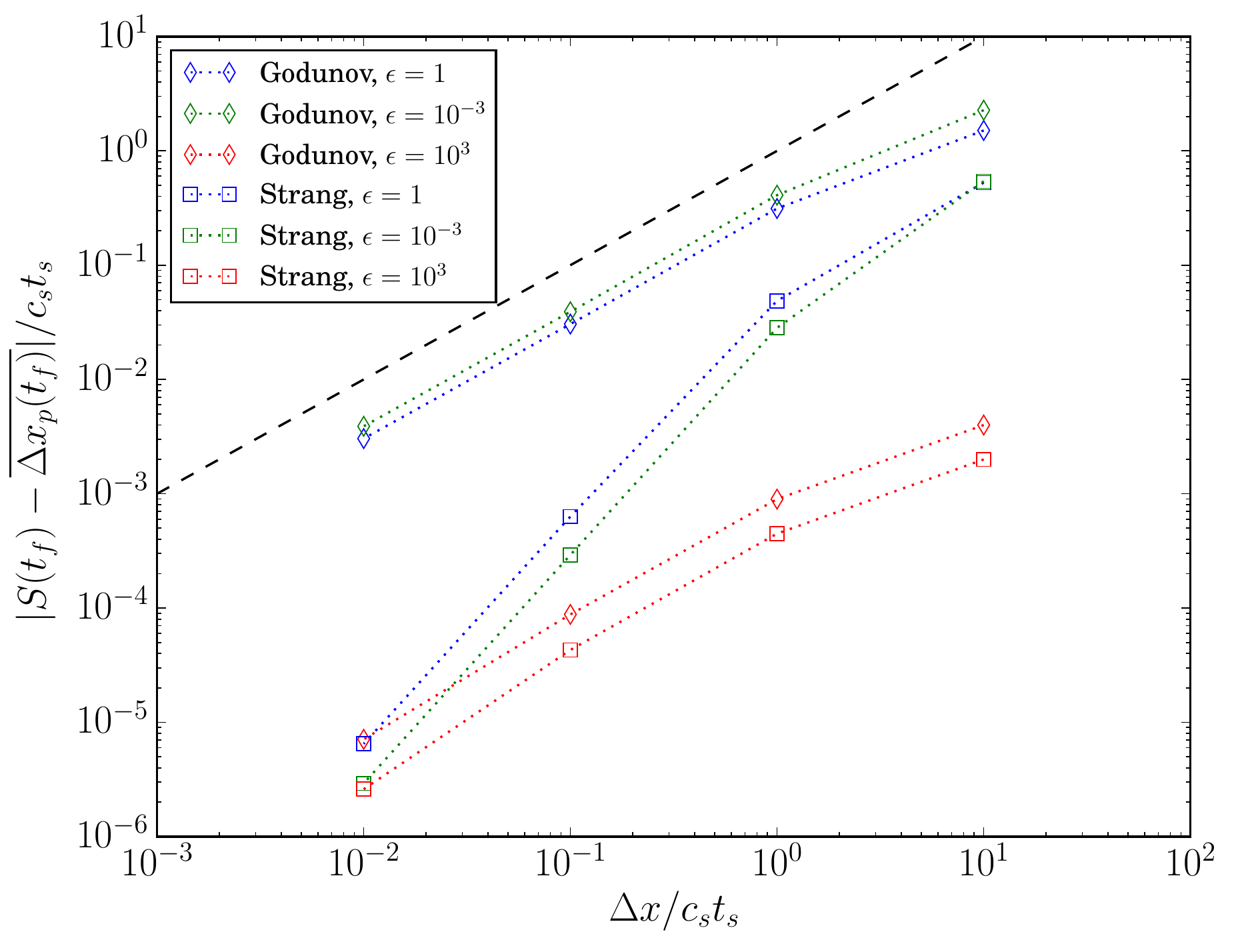}
\caption{Absolute error in the mean displacement of the particles as a function of cell size $\Delta x$ for the uniform streaming test.
The displacements are measured at a fixed final time of $t_f = 2t_s$ and are normalized by $c_s t_s$.
The diamond and the square symbols are the results using the Godunov and the Strang splitting methods, respectively.
The blue, the green, and the red colors denote a solid-to-gas density ratio of $\epsilon = 1$, $\epsilon = 10^{-3}$, and $\epsilon = 10^3$, respectively.
The dashed line indicates the cell size for a relative comparison of the errors.
Note that it takes exactly one time step to reach $t = t_f$ for $\Delta x = 10c_s t_s$.
Note also that the accuracy of the gas and the particle velocities we obtain in this test problem is close to machine precision irrespective of the resolution, as discussed in Section~\ref{SS:unistr}.}
\label{F:userr}
\end{center}
\end{figure}

Therefore, the only error remains to be considered is the displacement of the particles.
Similar to the velocities, the noise in the displacements of the particles is small and close to the machine-precision level, and thus the separation between adjacent particles remain highly constant.
Figure~\ref{F:userr} shows the absolute error in the mean displacement of the particles as a function of cell size $\Delta x$, and it can be seen that numerical convergence is achieved in a wide range of solid-to-gas density ratio $\epsilon$ with both the Godunov and the Strang splittings.
A few features can be observed from Figure~\ref{F:userr}.
Firstly, the accuracy is relatively insensitive to the solid-to-gas density ratio $\epsilon$ for $\epsilon \lesssim 1$, while it improves appreciably for $\epsilon \gg 1$.
Secondly, the Godunov splitting demonstrates the expected first-order convergence.
The Strang splitting, on the other hand, only achieves the second-order convergence when $\Delta x$ is sufficiently small such that the mutual drag timescale of $t_s / (1 + \epsilon)$ is resolved.
This behavior is similar to what is found in Section~\ref{SS:osc}.
Thirdly, the Strang splitting renders more accurate displacements when $\epsilon \lesssim 1$.
However, the difference between the Strang and the Godunov splittings significantly reduces when $\epsilon \gg 1$.
Finally, we note that in all cases, the errors are all below the cell size $\Delta x$, and hence the algorithm gives more precise displacements of the particles than what the resolution can provide for.

\section{TWO-DIMENSIONAL TESTS} \label{S:2d}
\subsection{Shear Waves} \label{SS:swav}

In the local-shearing-sheet approximation (see, e.g., equation~\eqref{E:gas_cont} or~\eqref{E:gas_mom}), the existence of the shear advection terms $-q\Omega x \partial f / \partial y$, where $f$ is any dynamical field variable, makes $e^{i\left[k_x(t)x + k_y y\right]}$ with a time-dependent $x$-wavenumber
\begin{equation}
k_x(t) = k_x(0) + q\Omega t k_y
\end{equation}
and a constant $y$-wavenumber $k_y$ a natural choice of the basis function.
This basis depicts a two-dimensional wave, in which the power in the azimuthal direction feeds the power in the radial direction, winding up the structure into a tighter and tighter spiral wave towards trailing morphology.
Substituting this basis into the two-fluid description of the particle-gas system without background radial gas pressure gradient (i.e., $a_x = 0$ in equation~\eqref{E:gas_mom}), \cite{YJ07} derived a set of ordinary differential equations for the (complex) amplitudes of the wave:
\begin{subequations} \label{E:swav}
\begin{align}
\tder{\hat{\rho}_g}{t} &=
    -\rho_{g,0}\left[i k_x(t)\hat{u}_x + i k_y\hat{u}_y\right],\\
\tder{\hat{u}_x}{t} &=
    2\Omega\hat{u}_y -
    \frac{\epsilon_0}{t_s}\left(\hat{u}_x - \hat{v}_x\right) -
    \frac{i k_x(t)c_s^2}{\rho_{g,0}}\hat{\rho}_g,\\
\tder{\hat{u}_y}{t} &=
    -(2 - q)\Omega\hat{u}_x -
    \frac{\epsilon_0}{t_s}\left(\hat{u}_y - \hat{v}_y\right) -
    \frac{i k_y c_s^2}{\rho_{g,0}}\hat{\rho}_g,\\
\tder{\hat{\rho}_p}{t} &=
    -\rho_{p,0}\left[i k_x(t)\hat{v}_x + i k_y\hat{v}_y\right],\\
\tder{\hat{v}_x}{t} &=
    2\Omega\hat{v}_y -
    \frac{1}{t_s}\left(\hat{v}_x - \hat{u}_x\right),\\
\tder{\hat{v}_y}{t} &=
    -(2 - q)\Omega\hat{v}_x -
    \frac{1}{t_s}\left(\hat{v}_y - \hat{u}_y\right),
\end{align}
\end{subequations}%
where $\rho_{g,0}$ and $\rho_{p,0}$ are the background uniform densities of the gas and the particles, respectively, $\epsilon_0 \equiv \rho_{p,0} / \rho_{g,0}$, and $c_s$ is the isothermal speed of sound.
This system of equations can be readily integrated numerically, and its solution serves as a convenient analytical benchmark to validate our implementation of the sheared periodic boundary conditions as well as the mutual drag force.

Applying our algorithm to this shear-wave test, we evolve the particle-gas system for a square domain $L \times L$ in the $xy$-plane with a single mode $k_x(0) = -2\pi / L$ and $k_y = 2\pi / L$.
We use a 64$\times$64 grid and a shear parameter of $q = 3/2$, and allocate one particle per cell.
The initial conditions are such that $\hat{\rho}_g(0) = \hat{u}_x(0) = \hat{\rho}_p(0) = \hat{v}_x(0) = \hat{v}_y(0) = 0$ while $\hat{u}_y(0) = 10^{-3}c_s$.
Note that in this case, it takes 22$\Omega^{-1}$ for the $x$-wavenumber to reach the Nyquist frequency, when the numerical diffusion and/or aliasing becomes significant.
In what follows, we vary the values of $\tau_s$ and $\epsilon_0$ to assess the performance of our algorithm, where $\tau_s \equiv \Omega t_s$.
We only present the results with the Godunov splitting method and note that the Strang splitting only slightly improves the accuracy.

\begin{figure}
\begin{center}
\plotone{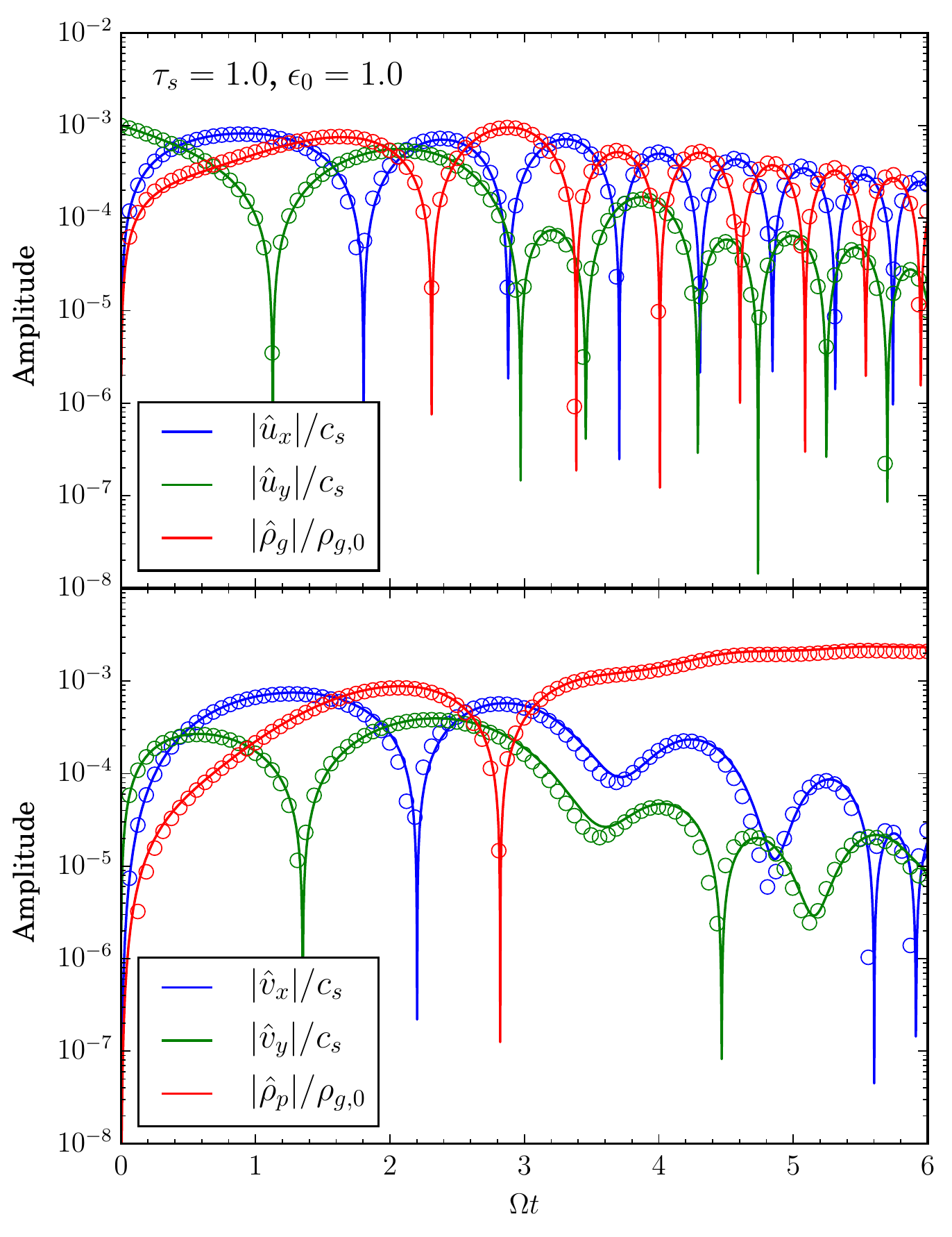}
\caption{Field amplitudes of a particle-gas shear wave as a function of time.
In this case, the system has a dimensionless stopping time of $\tau_s = 1$ and a solid-to-gas density ratio of $\epsilon_0 = 1$, and its initial conditions are described in Section~\ref{SS:swav}.
The solid lines are the analytical solutions obtained from integrating equations~\eqref{E:swav}, while the circles are the measurements from each and every time step of the simulation data obtained by our algorithm.
The top panel shows the velocity and density fields of the gas, while the bottom panel shows those of the particles.}
\label{F:swav1}
\end{center}
\end{figure}

First we consider the case of $\tau_s = 1$ and $\epsilon_0 = 1$, which makes the test exactly the same as was done in \cite{YJ07}.
Figure~\ref{F:swav1} shows the comparison between the analytical solutions obtained from integrating equations~\eqref{E:swav} and the measurements of the amplitudes on the simulation data obtained with our algorithm.
All the amplitudes of the shear wave from the simulation agree well with the analytical solutions for several orbital periods, with some minor deviation in the velocity field of the particles for $t \gtrsim 3.5\Omega^{-1}$.
Note that the time steps used here are significantly larger than those used in \cite{YJ07}.

\begin{figure}
\begin{center}
\plotone{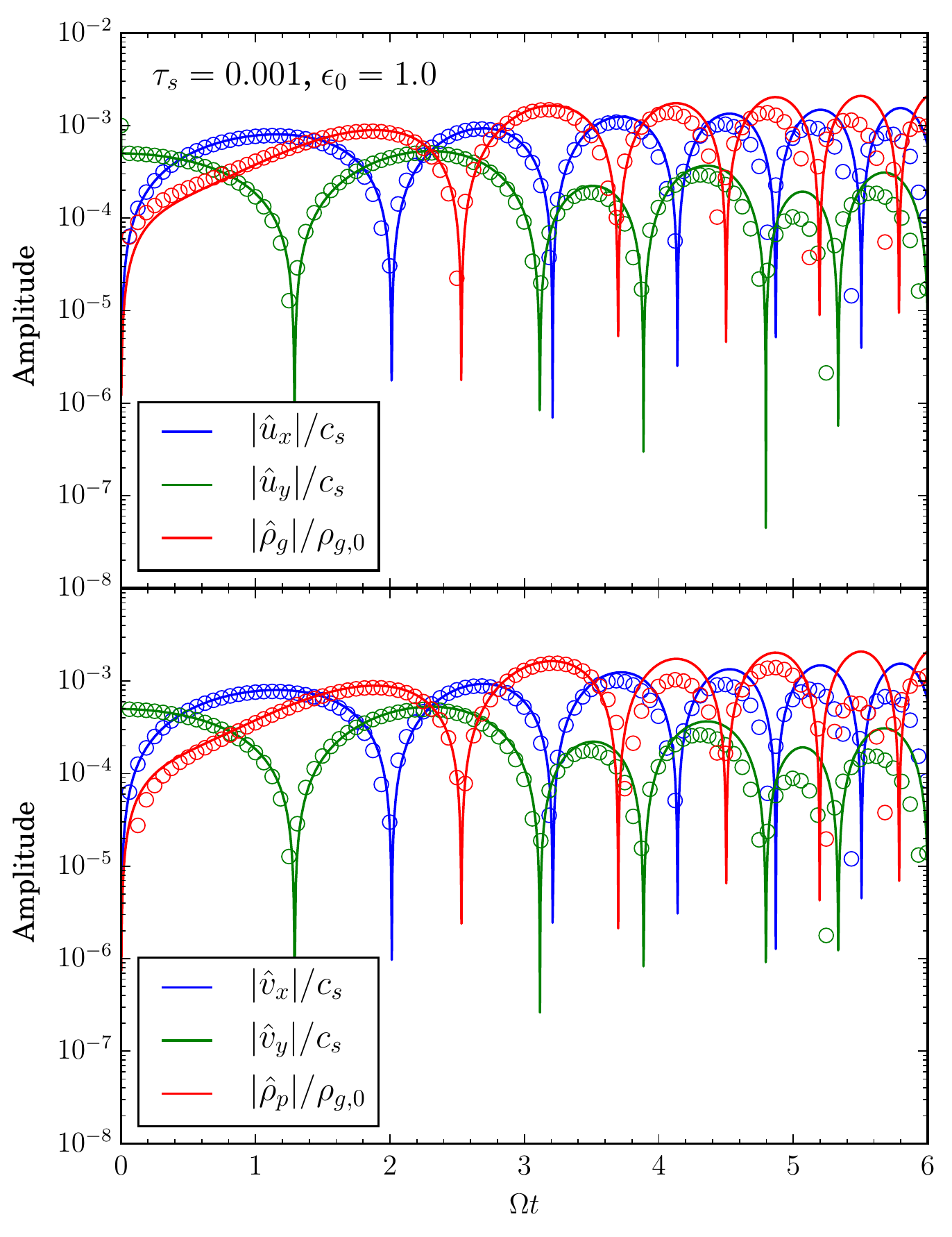}
\caption{Same as Figure~\ref{F:swav1}, except that $\tau_s = 10^{-3}$ and $\epsilon_0 = 1$.
Note that there exists an initial jump in $\hat{u}_y$ and $\hat{v}_y$ in the analytical solution.}
\label{F:swav2}
\end{center}
\end{figure}

Next we probe the case of small particles with $\tau_s = 10^{-3}$ and $\epsilon_0 = 1$, the result of which is shown in Figure~\ref{F:swav2}.
In this case, there exists an initial abrupt jump in the $u_y$ and the $v_y$ fields, followed by a smooth oscillatory evolution in the amplitudes as in the previous case.
An integration with the explicit method would require an extremely small time step to resolve and accurately capture this initial jump; it is not even numerically stable if the time step is larger than the width of this feature.
Our algorithm, in contrast, accurately predicts the velocity fields with a time step much longer than the timescale of this feature.
In spite of some initial minor deviation of the density fields, the solutions with our algorithm closely follow the analytical ones up to $t \sim 3.5\Omega^{-1}$, after which some noticeable deviation appears in both the gas and the particle fields.

\begin{figure}
\begin{center}
\plotone{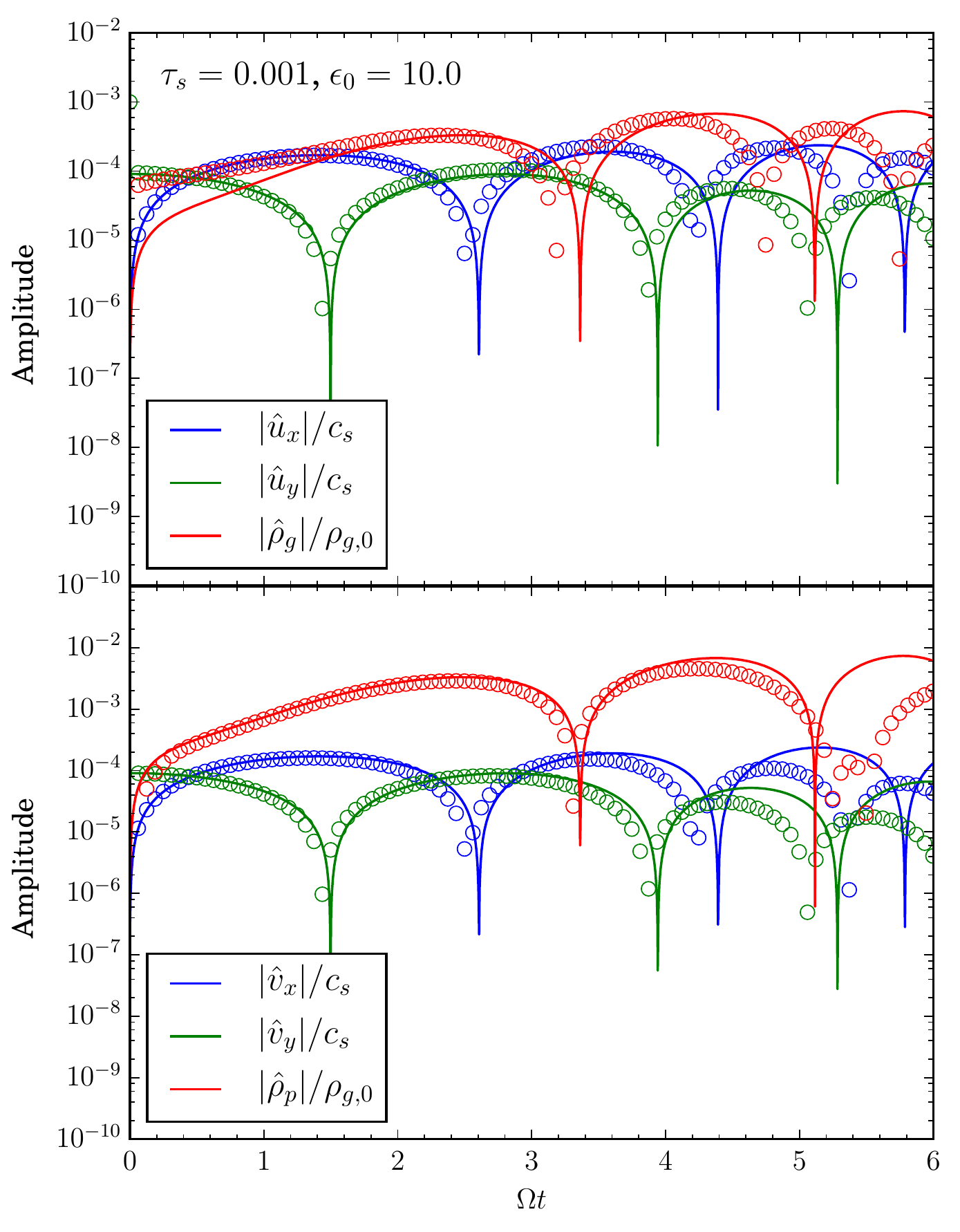}
\caption{Same as Figure~\ref{F:swav1}, except that $\tau_s = 10^{-3}$ and $\epsilon_0 = 10$.
Note that there exists an initial jump in $\hat{u}_y$ and $\hat{v}_y$ in the analytical solution.}
\label{F:swav3}
\end{center}
\end{figure}

Finally, we barge into the solid-dominated regime with $\tau_s = 10^{-3}$ and $\epsilon_0 = 10$, as shown in Figure~\ref{F:swav3}.
Similar to the previous case, a yet larger initial jump occurs in the $u_y$ field but a lesser one in the $v_y$ fields, and our algorithm accurately finds the first velocity fields with one long time step.
Although the algorithm also captures the density field of the particles relatively accurately, a significant error exists in the density field of the gas at the very first time step.
This error affects the accuracy of the subsequent evolution of the shear wave, the most prominent of which is in the frequency of the oscillation in the amplitude of each field.
Nevertheless, the general evolution of this shear wave is still reproduced by our algorithm up to $t \sim 4\Omega^{-1}$, where especially noticeable are the maximum amplitudes achieved in each oscillation of the fields.

\begin{figure}
\begin{center}
\plotone{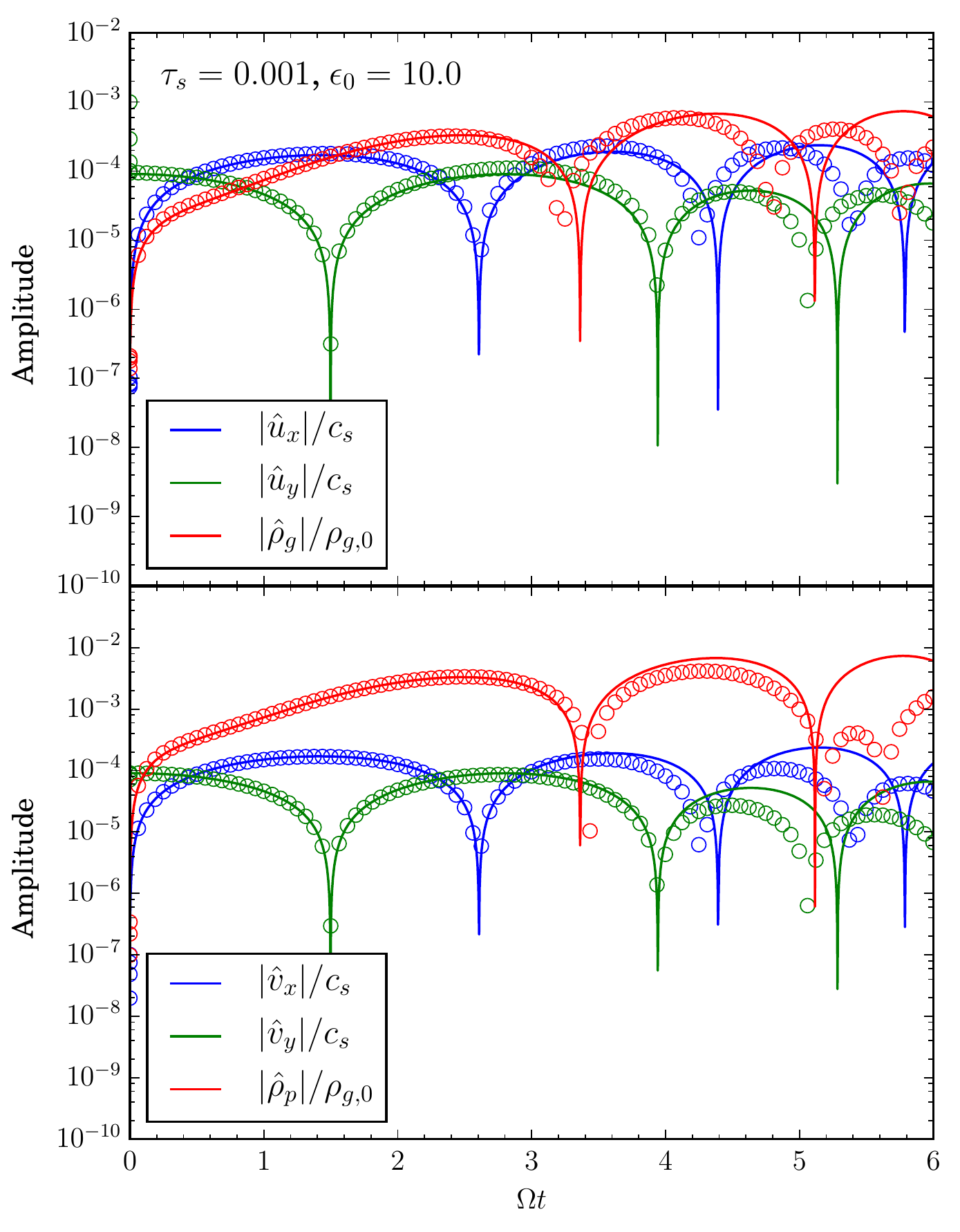}
\caption{Same as Figure~\ref{F:swav3}, except that the initial transient jump is resolved by four time steps.}
\label{F:swav4}
\end{center}
\end{figure}

This last case highlights that some inaccuracy in density fields can occur when there exists unresolved transient behavior in velocity, a situation reminiscent of the sedimentation benchmark presented in Section~\ref{SS:osc}.
The magnitude of this numerical error depends on how strong the change in velocity is, and in this test problem, increases with increasing background solid-to-gas density ratio.
Note that this kind of transient behavior often stems from initial conditions which are significantly out of equilibrium, as in this case, or impulses imposed onto the system in its course of evolution.
Once the impulse subsides and the system resumes smooth evolution, i.e., one that is resolved by the numerical time steps, our algorithm should exhibit high degree of accuracy, as demonstrated in the first case (or even the second case) of this section as well as other benchmarks presented in this work.
On the other hand, if the transients are of importance, one can always restrict the time steps so that those can be accurately captured.
As demonstrated in Figure~\ref{F:swav4}, simply resolving the initial transient jump by four time steps makes the evolution of all fields accurate up to $t \sim 3.2\Omega^{-1}$, a similar performance achieved in the earlier cases of Figures~\ref{F:swav1} and~\ref{F:swav2}.

\subsection{The Streaming Instability}

An important discovery in the theory of planet formation was that of the streaming instability by \cite{YG05}.
This instability efficiently concentrates centimeter/meter-sized solid particles in a protoplanetary gas disk to trigger gravitational collapse and form kilometer-sized planetesimals, circumventing the problematic fast radial drift of the solid particles \citep[e.g.,][]{JO07}.
The mutual drag force between the gas and the solid particles moving in a differentially rotating disk is an essential ingredient in this instability, and hence the results of the analysis carried out by \cite{YG05}, serendipitously, can be used as a rigorous touchstone to validate any numerical algorithm concerning the integration of this kind of the particle-gas system.
This is the goal of this section.

\subsubsection{Linear Modes} \label{SS:lsi}
\begin{figure*}[p]
\begin{center}
\epsscale{0.8}
\plotone{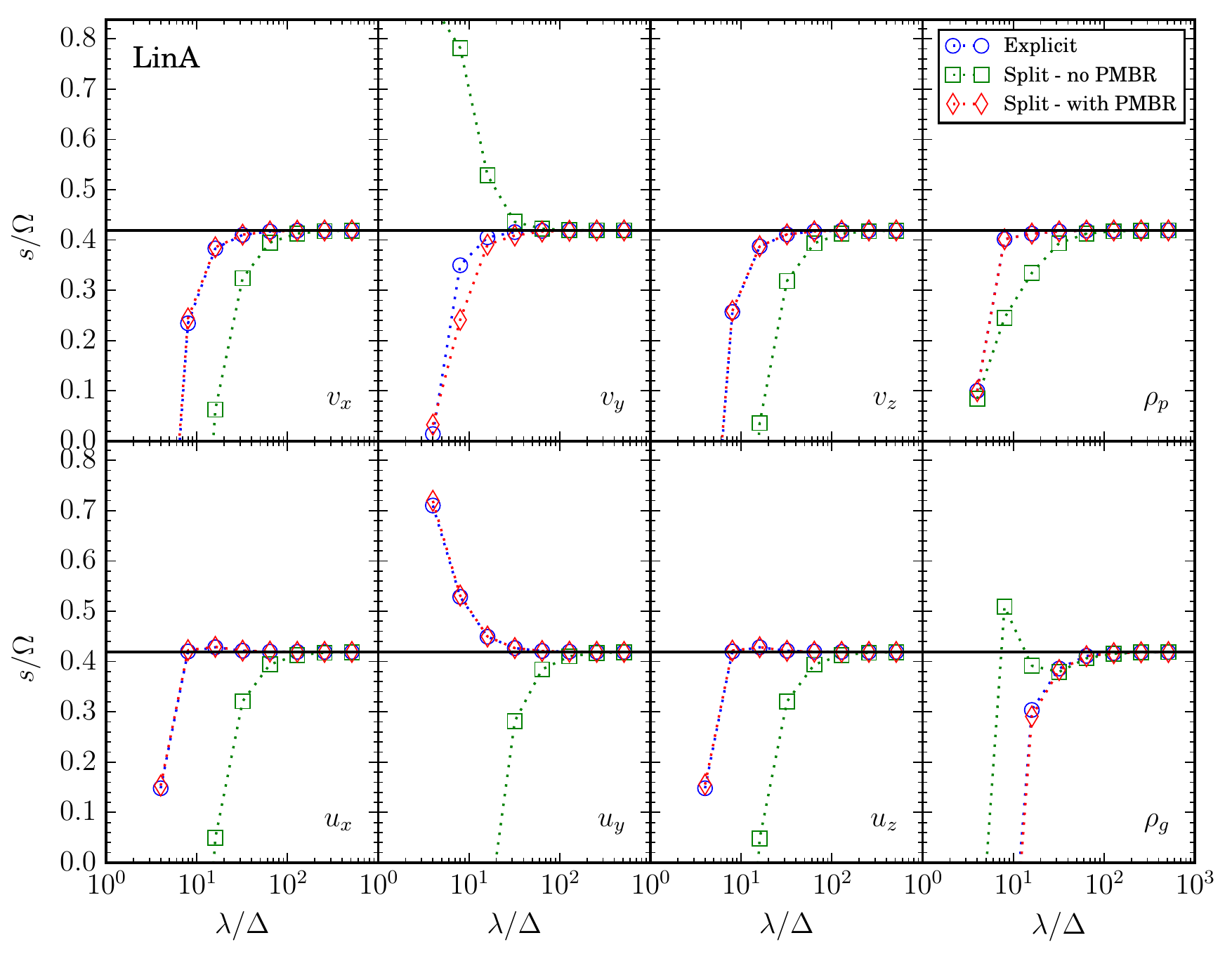}
\epsscale{1}
\caption{Measured growth rates as a function of resolution for the linA mode of the streaming instability from \cite{YJ07}.
This mode is for particles of dimensionless stopping time $\tau_s = 0.1$ and a background solid-to-gas density ratio of $\epsilon_0 = 3$.
The resolution is expressed in number of cells per wavelength $\lambda$ of the mode, where $\Delta$ is the cell size.
The rates $s$ are expressed in terms of the local angular speed $\Omega$ and are measured over the time interval $0 \le t \le 0.2P$, where $P = 2\pi / \Omega$ is the local orbital period.
The top row shows the three components of the particle velocity field $\vec{v}$ and the particle density field $\rho_p$, while the bottom row shows the three components of the gas velocity field $\vec{u}$ and the gas density field $\rho_g$.
The solid lines denote the theoretical growth rate.
The blue circles are the results obtained with the original explicit integration, while the red diamonds and the green squares are the results computed by our algorithm with and without the particle-mesh treatment of the back-reaction, respectively, which is described in Section~\ref{SS:upgas}.}
\label{F:linA}
\end{center}
\end{figure*}
\begin{figure*}[p]
\begin{center}
\epsscale{0.8}
\plotone{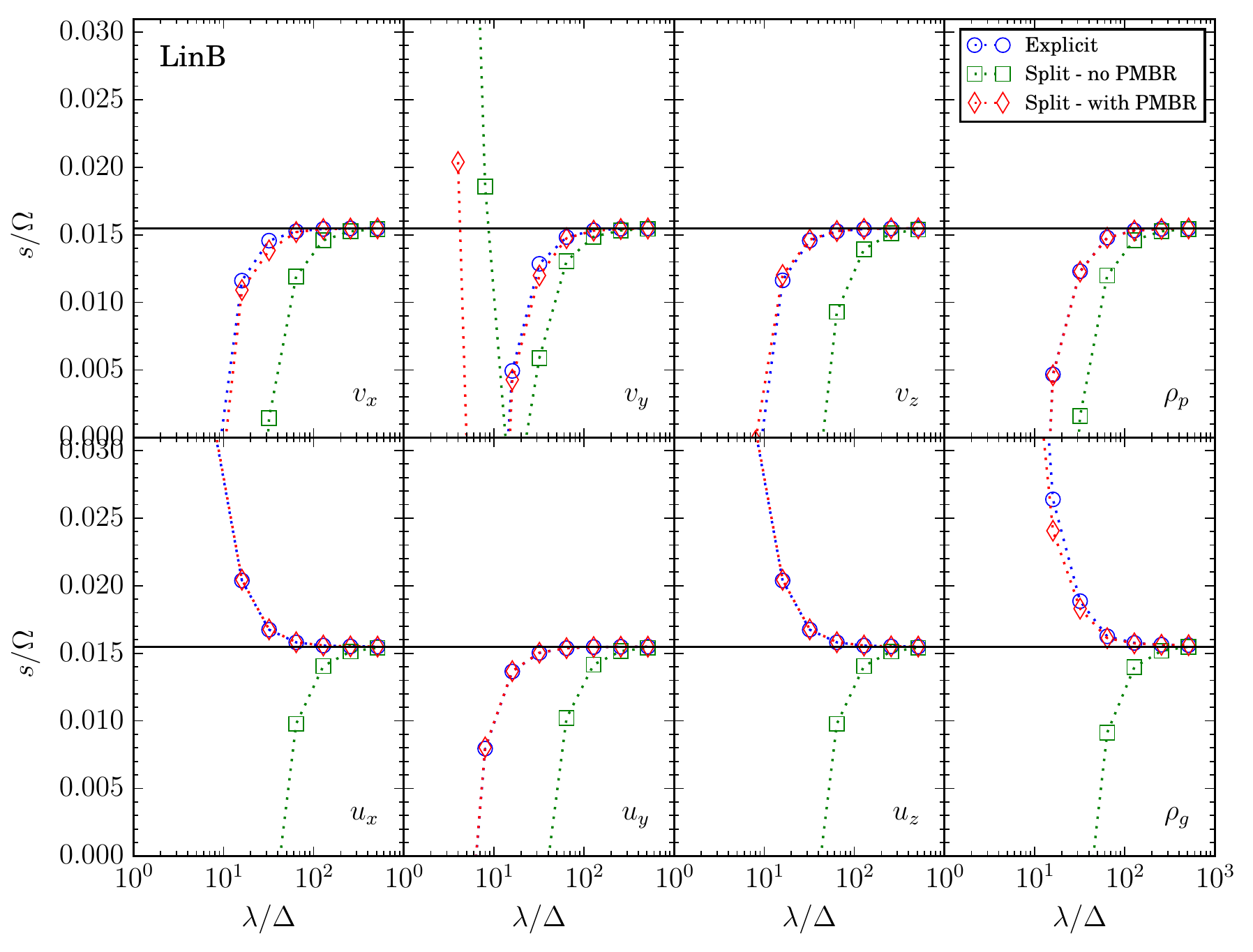}
\epsscale{1}
\caption{Measured growth rates as a function of resolution for the linB mode of the streaming instability from \cite{YJ07}.
This mode is for particles of dimensionless stopping time $\tau_s = 0.1$ and a background solid-to-gas density ratio of $\epsilon_0 = 0.2$.
The arrangement of the panels and the line styles are the same as those in Figure~\ref{F:linA}.
The growth rates are measured over the time interval $0 \le t \le P$.}
\label{F:linB}
\end{center}
\end{figure*}
\begin{figure*}[p]
\begin{center}
\epsscale{0.8}
\plotone{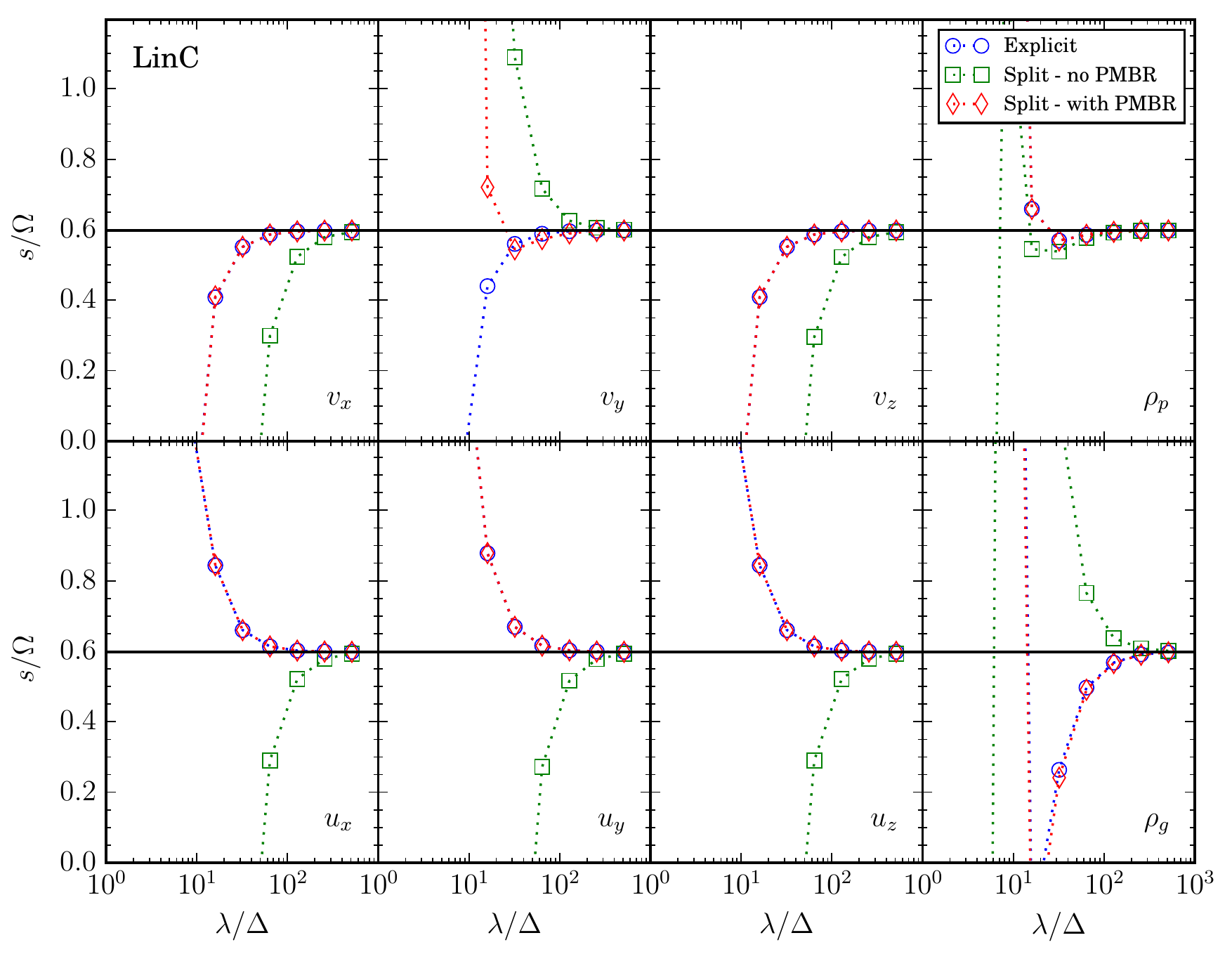}
\epsscale{1}
\caption{Measured growth rates as a function of resolution for the linC mode of the streaming instability from \cite{BS10b}.
This mode is for particles of dimensionless stopping time $\tau_s = 10^{-2}$ and a background solid-to-gas density ratio of $\epsilon_0 = 2$.
The arrangement of the panels and the line styles are the same as those in Figure~\ref{F:linA}.
The growth rates are measured over the time interval $0 \le t \le 0.02P$.}
\label{F:linC}
\end{center}
\end{figure*}
\begin{figure*}[p]
\begin{center}
\epsscale{0.8}
\plotone{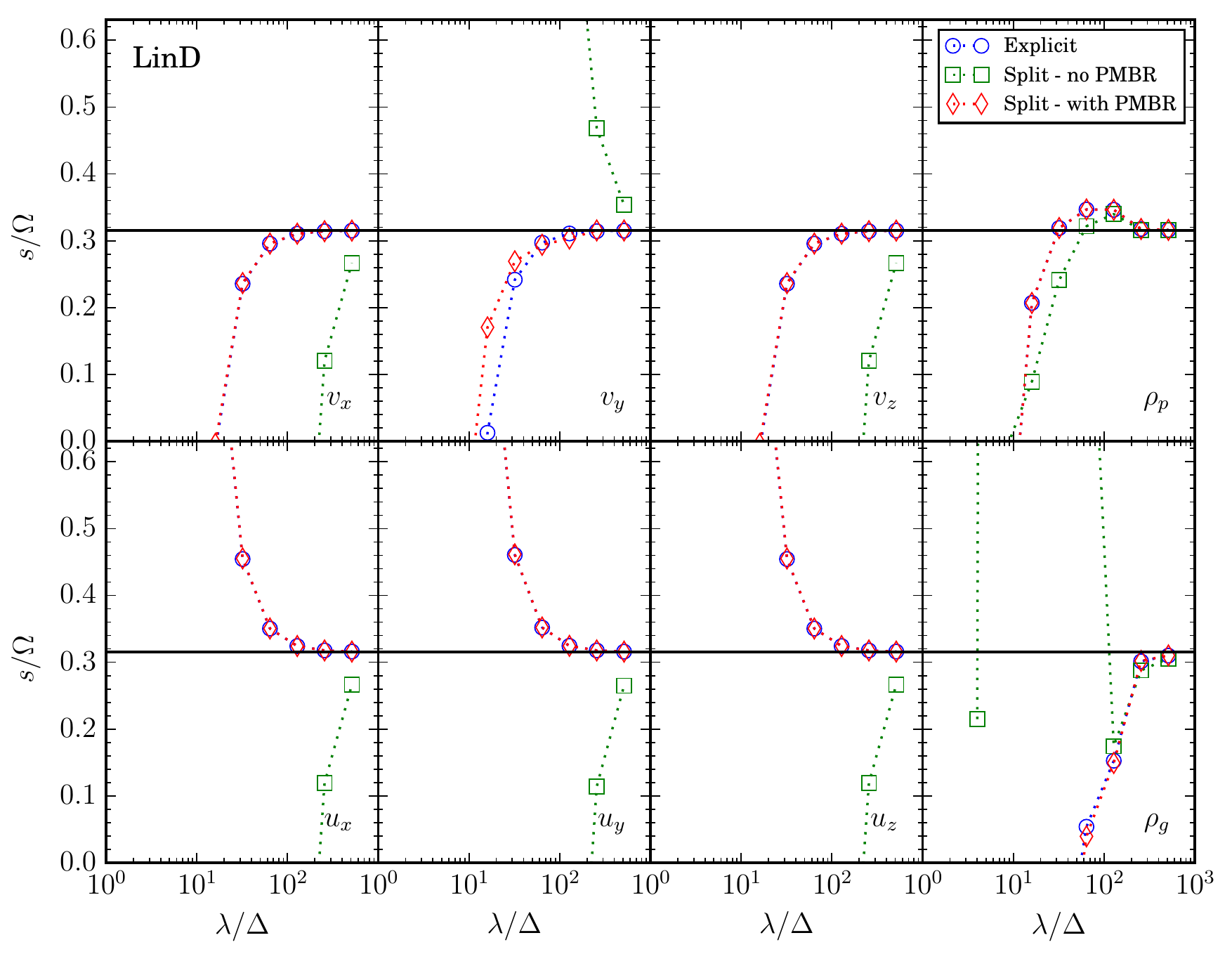}
\epsscale{1}
\caption{Measured growth rates as a function of resolution for the linD mode of the streaming instability from \cite{BS10b}.
This mode is for particles of dimensionless stopping time $\tau_s = 10^{-3}$ and a background solid-to-gas density ratio of $\epsilon_0 = 2$.
The arrangement of the panels and the line styles are the same as those in Figure~\ref{F:linA}.
The growth rates are measured over the time interval $0 \le t \le 0.01P$.}
\label{F:linD}
\end{center}
\end{figure*}

The streaming instability is a local, two-dimensional, axisymmetric instability with interpenetrating gas and particles under differential rotation.
Ignoring vertical gravity and assuming an isothermal equation of state for the gas, \cite{YG05} have performed linear analysis on this system and found the growth rate and wave speed of this instability and the corresponding eigenvector as a function of the wavenumber of the perturbation as well as all the relevant dimensionless parameters for the system.
The eigensystem was expressed in the frame of the center-of-mass velocity of the particle-gas system, which was not convenient for direct comparison with numerical simulations, and hence \cite{YJ07} have transformed these solutions back into the local-shearing-sheet frame and expressed them as a standing wave in the vertical direction while propagating in the horizontal direction.
The resulting eigenfunction is either even
\begin{align}
f_e(x,z) &= \left[\mathrm{Re}(\tilde{f})\cos\left(k_x x - \omega_R t\right)\right.
                -\nonumber\\&\qquad
            \left.\mathrm{Im}(\tilde{f})\sin\left(k_x x - \omega_R t\right)\right]
           e^{st}\cos k_z z,\label{E:lsie}
\end{align}
or odd
\begin{align}
f_o(x,z) &= -\left[\mathrm{Re}(\tilde{f})\sin\left(k_x x - \omega_R t\right)\right.
                +\nonumber\\&\qquad
             \left.\mathrm{Im}(\tilde{f})\cos\left(k_x x - \omega_R t\right)\right]
           e^{st}\sin k_z z,\label{E:lsio}
\end{align}
in the vertical direction, where $\tilde{f}$ is the complex amplitude, $\vec{k} = k_x\unitvec_x + k_z\unitvec_z$ is the wavenumber, and $\vec{\omega} = \omega_R + i s$ is the complex angular frequency.
The vertical component of the velocities of the gas and the particles assumes the odd parity, while all other dynamical fields assume the even parity.
On top of the background equilibrium state as given in equations~\eqref{E:uxeq}--\eqref{E:vyeq} and with an initially small amplitude for the perturbations, equations~\eqref{E:lsie} and~\eqref{E:lsio} then serve as both the initial conditions and the analytical solutions against which our algorithm is validated.

Four eigensystems have been published in the literature, all of which are the (nearly) fastest growing modes at the respective set of the dimensionless parameters.
We first test the linA and the linB modes in \cite{YJ07}.
They have the same dimensionless stopping time of $\tau_s = 0.1$ as well as the same background radial gas pressure gradient of $a_x = 0.1c_s\Omega$, where $c_s$ is the isothermal speed of sound and $\Omega$ is the local Keplerian angular frequency.
On the other hand, the linA mode has a background solid-to-gas density ratio of $\epsilon_0 = 3$, while the linB mode has $\epsilon_0 = 0.2$.
In our tests, the computational domain is such that one wavelength of the mode is fit in each direction.
We use one particle per cell, and use the method described in Appendix~C of \cite{YJ07} to seed a perturbation of $\tilde{\rho}_p = 10^{-6}$ in the density field of the particles.
Given their positions, the initial velocities of the particles are then set by the perturbation equations~\eqref{E:lsie} and~\eqref{E:lsio} as well as the equilibrium drift velocity equations~\eqref{E:vxeq} and~\eqref{E:vyeq}.
It is trivial to set the initial density and velocity field of the gas using equations~\eqref{E:lsie} and~\eqref{E:lsio} as well as equations~\eqref{E:uxeq} and~\eqref{E:uyeq}.
We measure the complex amplitude of the Fourier mode $(k_x, k_z)$ as a function of time in our simulation data, and then use the linear regression on the magnitude of the amplitude to find the (exponential) growth rate $s$.
We vary the resolution to seek the convergence of our measured rate as compared to the analytical one.
For comparison purposes, we use three different integration schemes for this test.
One is the explicit integration, as originally used by \cite{YJ07}.
The other two are our algorithm either with or without particle-mesh back-reaction (PMBR) to the gas velocity field, as described in Section~\ref{SS:upgas}, and we only present the results with the Godunov splitting here.

\begin{figure*}
\begin{center}
\plotone{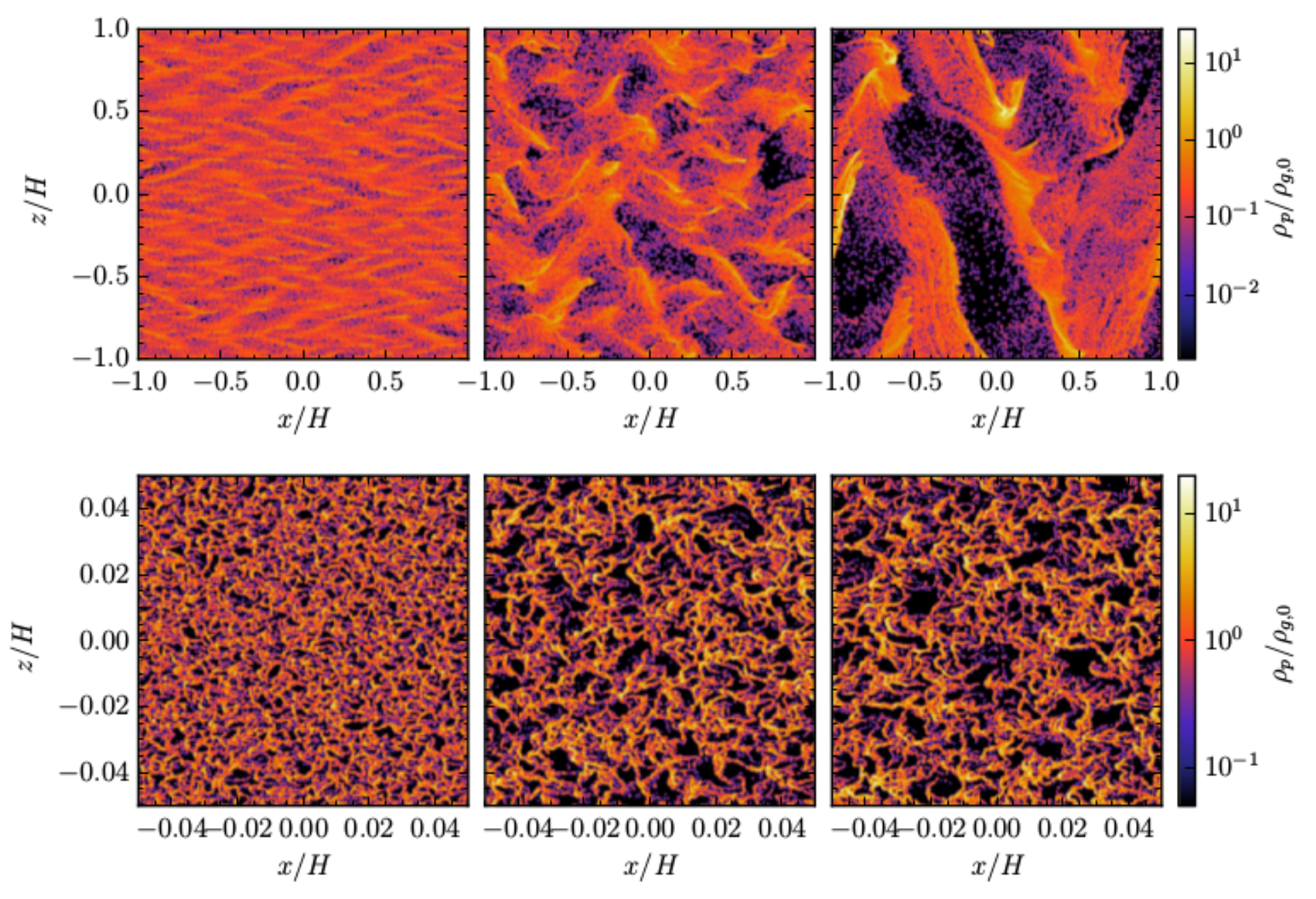}
\caption{Nonlinear evolution of the streaming instability computed by our algorithm.
The top and the bottom rows show the models~BA and~AB, respectively, taken from \cite{JY07}.
The models are axisymmetric and in the $xz$-plane, where the radial axis $x$ and the vertical axis $z$ are measured in terms of the gas scale height $H$.
The color images depict the density field of the particles $\rho_p$ in terms of the initial uniform gas density $\rho_{g,0}$.
The top panels show from left to right the snapshots at $t = 4P$, $14P$, and $50P$, respectively, while the bottom panels show the snapshots at $t = 0.4P$, $1.4P$, and $5P$, respectively, where $P$ is the local orbital period.}
\label{F:nsi}
\end{center}
\end{figure*}

The results are shown in Figures~\ref{F:linA} and~\ref{F:linB}.
As can be seen, the growth rates for all three schemes converge to the theoretical one with resolution.
For the linA mode, our algorithm \emph{with PMBR} achieves virtually the same performance of the explicit integration.
While it requires a resolution of around 32--64~points per wavelength $\lambda$ for the growth rate of the $\rho_g$ field to approach within 5\% of the theoretical one, it requires around 16--32~points per $\lambda$ for those of the $v_x$, $v_z$, and $u_y$ fields, around 16~points per $\lambda$ for that of the $v_y$ field, and only around 4--8~points per $\lambda$ for that of the $\rho_p$, $u_x$, and $u_z$ fields.
Albeit convergent, our algorithm \emph{without PMBR} has a much slower convergence rate, requiring a resolution of around 64--128~points per $\lambda$ for the growth rates of most fields, except around 32--64~points per $\lambda$ for those of the $\rho_p$ and $\rho_g$ fields and around 16--32~points per $\lambda$ for that of the $v_y$ field.
As for the linB mode, our algorithm \emph{with PMBR} has again similar growth rates as achieved by the explicit integration.
A resolution of around 32--64~points per $\lambda$ is required, except for the $u_y$ field, which requires only a resolution of around 16--32~points per $\lambda$.
Our algorithm \emph{without PMBR} remains inferior, requiring a resolution of around 128--256 points per $\lambda$ in all fields except for the $v_y$ field, which requires a resolution of around 64--128 points per $\lambda$.

The other two modes are the linC and the linD modes from \cite{BS10b}.
These modes push the limit of the dimensionless stopping time down to $\tau_s = 0.01$ and $\tau_s = 0.001$, respectively.
They have the same solid-to-gas density ratio of $\epsilon_0 = 2$, as well as the same background radial gas pressure gradient $a_x = 0.1c_s\Omega$ as do the linA and the linB modes.
Our test results are shown in Figures~\ref{F:linC} and~\ref{F:linD}.
Again, the growth rates for all three schemes demonstrate convergence towards the theoretical one.
Moreover, the performance of our algorithm \emph{with PMBR} remains essentially the same as that of the explicit integration for all fields.
For the LinC mode, a resolution of around 32--64~points per wavelength $\lambda$ is required for all the velocity fields, while a resolution of around 16--32~points per $\lambda$ and that of around 128--256~points per $\lambda$ are required for the $\rho_p$ and the $\rho_g$ fields, respectively.
For the linD mode, a resolution of around 64--128~points per $\lambda$ is required for all the velocity fields, while that of around 128--256~points per $\lambda$ is required for both the density fields.
We note that the performance of these integrators for these modes is relatively better in comparison with that performed in \cite{BS10b}.
As for our algorithm \emph{without PMBR}, the convergence is again much slower as compared to the other two schemes, and it does not even converge to within 5\% of the theoretical rate at a resolution of 512~points per $\lambda$ for the linD mode except for the density fields, which requires a resolution of around 256~points per $\lambda$.

These results demonstrate the excellent performance of our algorithm in reproducing the linear modes of the streaming instability, especially with a wide range of the dimensionless stopping time $\tau_s$.
Moreover, they justify the necessity of the implementation of PMBR to the gas velocity in our algorithm as described in Section~\ref{SS:upgas}.
We note here that there exists no noticeable improvement in the test results when using our algorithm with the Strang splitting method.
In the next subsection, we continue to explore our algorithm in the nonlinear saturation of the streaming instability.

\subsubsection{Nonlinear Saturation}
In a companion paper to \cite{YJ07}, \cite{JY07} investigated the nonlinear saturation of the streaming instability and found two distinctive patterns that could develop in this stage.
One is for marginally coupled solid particles with $\tau_s = 1$, while the other is for strongly coupled solid particles with $\tau_s = 0.1$.
The $\tau_s = 1$ particles initially concentrate themselves into short stripes that are close to horizontal but tilt alternately in the vertical direction with a separation consistent with the wavelength of the fastest growing mode of the streaming instability.
These stripes then undergo inverse cascade and merge themselves into larger and larger stripes that tilt more and more towards the vertical direction.
In the fully saturated stage, long, strongly concentrated filaments with a wide separation float either upwards or downwards with much reduced radial drift motion.
On the other hand, the $\tau_s = 0.1$ particles initially create numerous small voids by random, locally divergent motions.
These voids then undergo inverse cascade and merge themselves into larger and larger voids, driving the particles into their rims.
In the fully saturated stage, voids of various sizes move around in random directions, while the particles stream along the alleyways in between these voids.

Using our algorithm along with the Godunov splitting method, we rerun Model~BA and Model~AB in \cite{JY07}.
Model~BA contains marginally coupled particles with $\tau_s = 1$ and has a solid-to-gas density ratio of $\epsilon_0 = 0.2$, while Model~AB contains strongly coupled particles with $\tau_s = 0.1$ and has a solid-to-gas density ratio of $\epsilon_0 = 1$.
Both models have a background radial gas pressure gradient of $a_x = 0.1c_s\Omega$.
Adopting the same computational domains and resolutions as used in \cite{JY07}, we construct a \twodim{256}{256}~grid with a domain of \twodim{2$H$}{2$H$} and \twodim{0.1$H$}{0.1$H$} for Model~BA and Model~AB, respectively, where $H$ is the scale height of the gas.
We allocate on average one particle per cell, which is sufficient in capturing the overall density distribution function of particles at a fixed resolution, as has been shown by \cite{BS10b}.
The initial velocities of the gas and the particles are those under mutual drag equilibrium, equations~\eqref{E:uxeq}--\eqref{E:vyeq}.
The initial density field of the gas is uniform, while the initial positions of the particles are random to seed the streaming instability at all scales.
The resulting evolutions of the density field of the particles are shown in Figure~\ref{F:nsi}.
As can be seen, we have reproduced the two aforementioned distinct patterns in the saturated state of the streaming instability, with quantitative similarity as the same models in \cite{JY07}.
This again illustrates the consistency between the original explicit integration and our algorithm in this work.

\section{THREE-DIMENSIONAL TEST} \label{S:3d}

Finally, we explore our algorithm with its full three-dimensional glory.
To the best of our knowledge, there exists no appropriate analytical model that contains all the physics considered in our algorithm.
We therefore resort to our earlier nonlinear simulations published in \cite{YJ14} as our base for comparison, which employed the technique of explicit integration.

\subsection{Sedimentation and Streaming Turbulence} \label{SS:sed}
\begin{figure*}
\begin{center}
\plotone{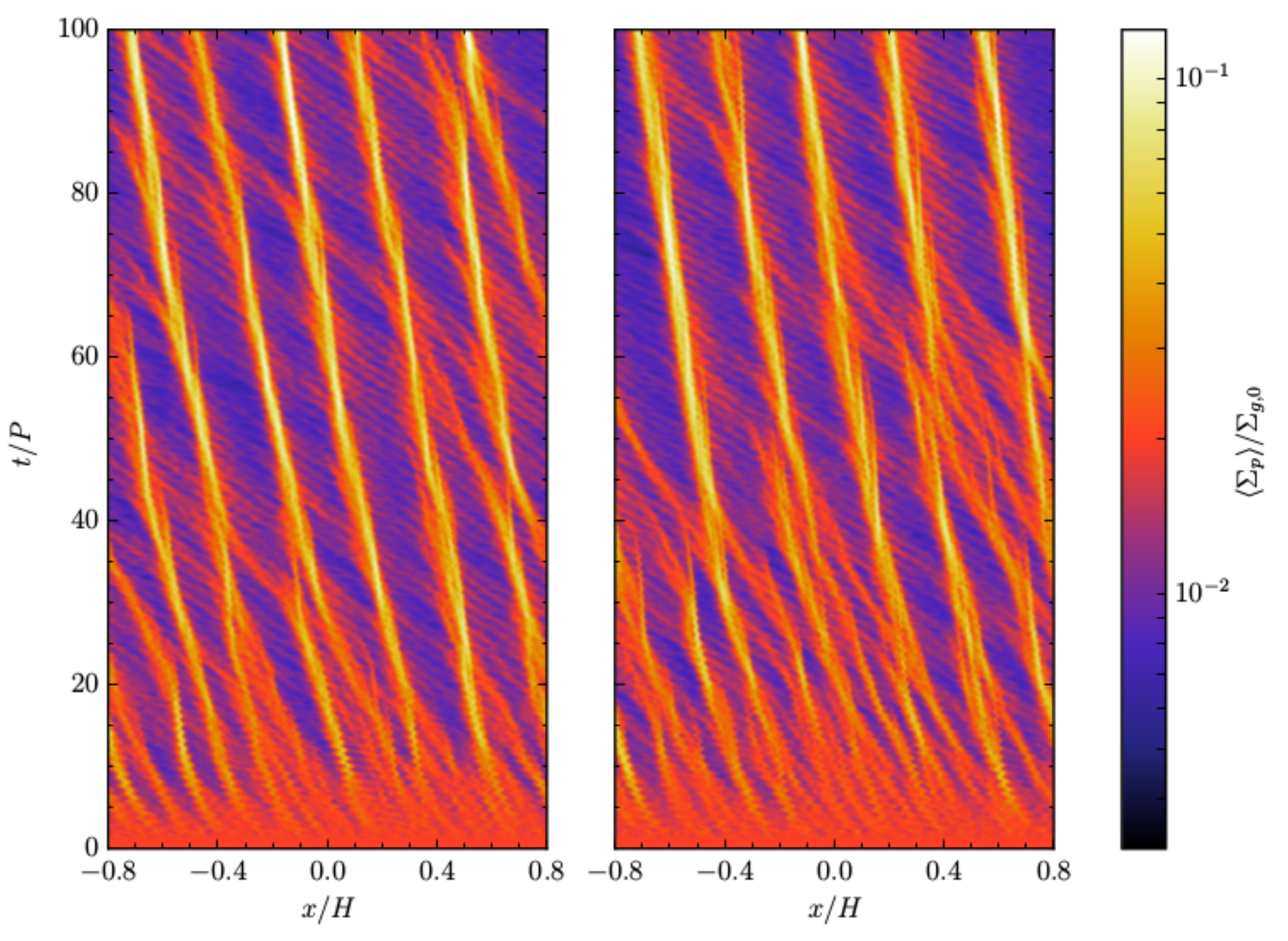}
\caption{Comparison of the azimuthally-averaged column density of solids as a function of radial position and time.
The column density of solids $\Sigma_p$ is normalized by the initial gas column density $\Sigma_{g,0}$, while the radial position $x$ and the time $t$ are normalized by the gas scale height $H$ and the local orbital period $P$, respectively.
The left panel shows the simulation with the original explicit integration in \cite{YJ14}, while the right panel shows that with our algorithm in this work.
The models are otherwise the same, with a computational domain of \threedim{1.6$H$}{1.6$H$}{0.2$H$} and a resolution of 160~points per $H$.}
\label{F:sedsig}
\end{center}
\end{figure*}

In \cite{YJ14}, we systematically studied the dependence of the streaming turbulence in a sedimented layer of solid particles on the computational domain of the simulations.
With particles of dimensionless stopping time $\tau_s \simeq 0.31$ and a background gas pressure gradient of $a_x = 0.1c_s\Omega$, where $c_s$ is the isothermal speed of sound and $\Omega$ is the local Keplerian angular frequency, we found that multiple radial filamentary concentrations of solids can be driven by the streaming instability and that the characteristic separation between adjacent filaments is on the order of 0.2$H$, with $H$ being the vertical scale height of the gas.

Using our algorithm along with the Godunov splitting method, we rerun otherwise exactly the same model in \cite{YJ14} that has a computational domain of \threedim{1.6$H$}{1.6$H$}{0.2$H$} and a resolution of 160~points per $H$.
Figure~\ref{F:sedsig} compares the resulting radial concentrations of solids between the simulations in \cite{YJ14} and in this work.
As can be seen in the comparison, we obtain excellent agreement between the two simulations.
Despite minor differences in the stochastic erosion and accretion of solids from and onto the filaments, the total number of filaments, their radial drifts, and the magnitude of the solid density are all quantitatively and evolutionarily similar.
This experiment demonstrates the consistency between the two techniques in three dimensions and yet again the robustness of our algorithm.

\begin{figure}
\begin{center}
\plotone{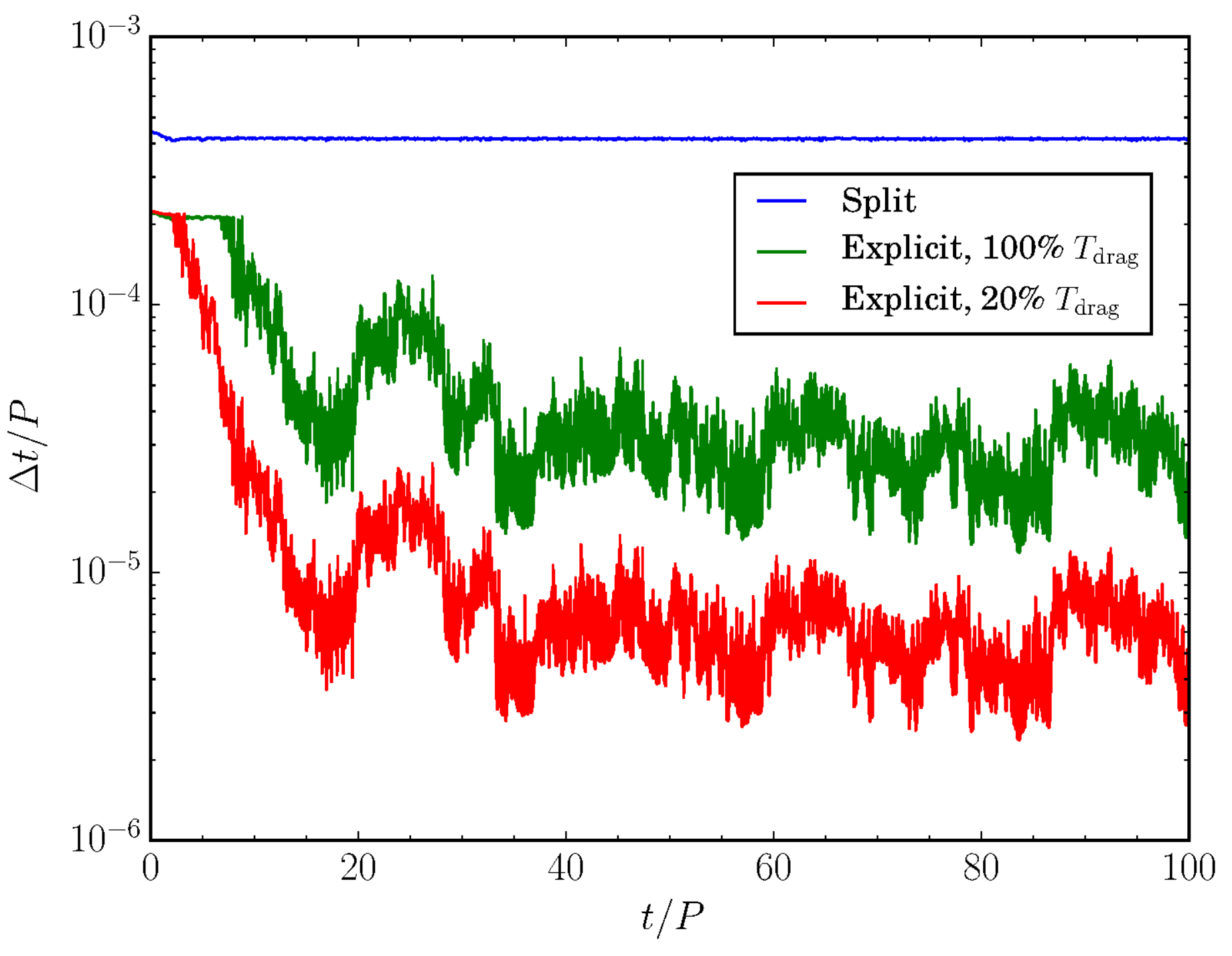}
\caption{Comparison of the time steps used in the simulations of Figure~\ref{F:sedsig} with either the explicit integration in \cite{YJ14} or our algorithm in this work.
Both the time step $\Delta t$ and the evolution time $t$ are normalized by the local orbital period $P$.
For the explicit integration, two possibilities are shown; one is limited by 100\% $T_\mathrm{drag}$, and the other is limited by 20\% $T_\mathrm{drag}$, where $T_\mathrm{drag}$ is the characteristic time of the mutual drag force between the gas and the particles.
See Section~\ref{SS:sed}.}
\label{F:seddt}
\end{center}
\end{figure}

In Figure~\ref{F:seddt}, we compare the time steps used in the two simulations.
For the simulation in \cite{YJ14}, the time steps were initially determined by the hydrodynamic Courant condition but soon dropped by roughly one order of magnitude beginning at $t \simeq 8P$, where $P$ is the local orbital period.
This is predominantly limited by the drag time $T_\mathrm{drag}$, which is the timescale for the exponential decay in the relative velocity between the gas and the particles due to their mutual drag interactions, as discussed in Section~\ref{S:intro}.
In the \textsc{Pencil Code}, it is approximated by $T_\mathrm{drag} \simeq (1 + \rho_g V / N_p m_p) t_s$, where $\rho_g$ is the gas density in the cell, $V$ is the volume of the cell, $N_p$ is the total number of (super-)particles in the cell, $m_p$ is the mass of a (super-)particle, and $t_s$ is the physical stopping time.
Note that we used 100\% of $T_\mathrm{drag}$ as our time-step limiter in \cite{YJ14},\footnote{In other words, we did not impose any artificial increase of the stopping time $t_s$ for cells with high local solid-to-gas density ratios; see Section~\ref{S:intro}.} which, although speeds up the simulations, could amount to a relative error of roughly 10\% in relative velocity in the worst-case scenario.
The \textsc{Pencil Code} defaults to use 20\% of $T_\mathrm{drag}$ as a time-step limiter to guarantee better accuracy, and thus we also show the time steps that would have been used in this case in Figure~\ref{F:seddt}.
As can be seen, the drag time begins to dominate earlier at $t \simeq 3P$ and the time steps drop by another order of magnitude than those used in \cite{YJ14}.

On the other hand, our algorithm is not limited by the drag time $T_\mathrm{drag}$ at all.
As shown in Figure~\ref{F:seddt}, the time steps remain almost constant and are only determined by the hydrodynamic Courant condition.\footnote{In \cite{YJ14}, we used a Courant number of 0.4, while in this work we use that of 0.8 with our algorithm.  Note that with the technique of explicit integration when the drag time $T_\mathrm{drag}$ dominates the time-step limiter, the Courant number used is irrelevant.}
These time steps are in drastic contrast to those used in the simulation of \cite{YJ14}, the latter of which are more than one order of magnitude smaller, and would have been more than two orders of magnitude smaller if 20\% of $T_\mathrm{drag}$ were adopted.
Therefore, this experiment also illustrates the exceptional efficiency of our algorithm, which we set out to achieve in this work.\footnote{As of this writing, our implementation of this algorithm in three dimensions requires roughly three times longer wallclock time \emph{per time step} than the explicit integration.  The benchmark was conducted on the Alarik supercomputing system at Lunarc in Lund University, Sweden.}

\section{CONCLUDING REMARKS} \label{S:conc}

In summary, we have devised an accurate, efficient numerical algorithm to directly integrate the mutual drag force in a system of Eulerian gas and Lagrangian solid particles.
Despite the entanglement between the gas and the particles due to the conventional particle-mesh construct, we have effectively decomposed the globally coupled system of equations for the mutual drag force and been able to integrate this system on a cell-by-cell basis.
Analytical solution exists for the temporal evolution of each cell, which we use to achieve the highest degree of accuracy.
This solution relieves the time-step constraint posed by the mutual drag force, making simulation models with small particles and/or strong local solid concentration significantly more amenable.
We have used an extensive suite of benchmarks with known solutions --- in one, two, and three dimensions --- to validate our algorithm and found satisfactory consistency in all cases.

Even though the Strang splitting is formally higher order, we find that its use with our algorithm does not offer significant advantage over the Godunov splitting, especially in multidimensional models.
In our one-dimensional benchmarks, both splittings predict virtually the same velocities, and hence they give the same accuracy in velocities and have the same behavior in numerical convergence.
As for the accuracy in particle positions, although the Strang splitting demonstrates expected second-order convergence for small time steps, it degrades to first-order convergence and does not give noticeably improved accuracy than the Godunov splitting for large time steps.
In our multidimensional benchmarks, on the other hand, no appreciable difference in either density or velocity fields for the gas and the particles exists between the two splitting methods.
We emphasize that since our objective is to relieve the time-step constraint due to the mutual drag force, the use of large time steps is of more interest here.
Moreover, note that since the Strang splitting requires one more run of either our algorithm or the other integrator than the Godunov splitting (see Appendix~\ref{S:os}), the former is significantly more expensive than the latter, let alone other operator splitting methods of even higher orders.
Therefore, we find that employing the Godunov splitting with our algorithm is sufficient and more economical for practical purposes.

The simulation models to which our algorithm can be applied are more general than what have been presented in this work.
First of all, our algorithm only concerns the mutual drag force as well as the rotation/shear-related source terms and the background accelerations.
All other physical processes are consolidated in the other half of the operator-split system of equations as in equations~\eqref{E:hypsys} and are integrated independently of our algorithm.
In this regard, the particle-gas system that can be considered using our algorithm is unconstrained and potentially arbitrary.
Moreover, our algorithm is not limited to the local-shearing-sheet approximation.
For instance, a non-rotating inertial frame, either rectangular or curvilinear, is simply a degenerate case of equations~\eqref{E:src} with the angular frequency $\Omega = 0$, and thus the same solution of Section~\ref{SS:asol} applies except that the equilibrium velocities (equations~\eqref{E:uxeq}--\eqref{E:vyeq}) need to be modified accordingly.
As for a rotating cylindrical coordinate system, the shear acceleration terms in equations~\eqref{E:src} are replaced by the centrifugal force as well as the external gravity, but an analytical counterpart of the solution in Section~\ref{SS:asol} can still be obtained from the modified system of equations.
Lastly, note also that it is not required for the particles to have the same mass, as shown in equation~\eqref{E:epsj}.

Some elaboration is needed when generalizing our procedure for a system with particles of various stopping times.
A closed-form analytical solution as that in Section~\ref{SS:asol} is only possible when the stopping time $t_s$ is a constant for all particles.
For the case of independent stopping time for each particle, nevertheless, it becomes merely a matter of adopting a numerical method that can accurately and efficiently approximate the solution of the corresponding system of equations~\eqref{E:src} in place of the analytical solution of Section~\ref{SS:asol}.
Given the stiffness of the mutual drag force, an implicit method is likely to be required for this purpose.
However, since we have effectively decoupled the system and made it possible to integrate it on a cell-by-cell basis, the individual integration task for each cell does not amount to an unmanageable size of matrix inversion, for instance.
Therefore, we expect that this extra complexity would not lead to noxious reduction in computational efficiency with our algorithm.

Having the time-step constraint due to stiff mutual drag force removed, our algorithm presented in this work will prove to be useful in the study of dust dynamics in protoplanetary disks.
Specifically, a model containing numerous mm/cm-sized pebbles and/or with high mass loading of solids can be simulated as efficiently as a model containing m-sized boulders without strong concentration of them.
In other words, it becomes feasible to evolve small solid particles interacting with a gas disk in relatively high numerical resolution and long simulation time.
This capability is particularly important in further study of the streaming instability with mm/cm-sized pebbles and its connection to planetesimal formation, as pioneered by \cite{CJD15}, since the wavelength of the fastest growing mode decreases with decreasing particle size and the corresponding growth rate is low when the initial solid abundance is low \citep{YG05,YJ07}.
For the same reasons, our algorithm may also find its use in investigating the dynamics of pebbles with the influence of turbulence, vortices, or other large-scale structures in the gas disk, and the corresponding observational consequences.

\acknowledgments
We thank James M.\ Stone for motivating this project.
We would also like to thank Shu-ichiro Inutsuka for his discussion on this work.
Significant improvement in the clarity of the manuscript was attributed to the referee's comments.
Part of the code development and the benchmarks presented in this work were performed on resources provided by the Swedish National Infrastructure for Computing (SNIC) at Lunarc in Lund University, Sweden.
This research was supported by the European Research Council under ERC Starting Grant agreement 278675-PEBBLE2PLANET.  A.~J.\ is grateful for financial support from the Knut and Alice Wallenberg Foundation and from the Swedish Research Council (grant 2010-3710).

\appendix
\section{NUMERICAL STRESSES INTRINSIC IN THE PARTICLE-MESH METHOD} \label{S:ns}

In this section, we demonstrate by a simple example that the mutual drag force between the gas and the particles leads to numerical stresses under the particle-mesh construct.
Consider a system of uniform gas and uniformly distributed particles.
Suppose that all the velocities of gas and particles align in the $y$-direction and depend on only $x$.
The evolution of this system then becomes a one-dimensional problem in the $x$-direction with all the velocities in the transverse direction, so the densities of both the gas and the particles remain uniform over time.
We construct a regular grid for the gas with one particle at each cell interface, i.e., the center of cell $k$ is at $x_k = (k - 1/2)\Delta x$ and the $j$-th particle is at $x_{p,j} = j \Delta x$, where $\Delta x$ is the cell size.
The system of equations for this problem reads, from Equations~\eqref{E:src} and~\eqref{E:epsj},
\begin{align}
\tder{u_k}{t} 
=& \frac{\epsilon}{t_s}\sum_j W(x_k - x_{p,j})(v_j - u_k),\label{E:nsdudt}\\
\tder{v_j}{t}
=& \frac{1}{t_s}\sum_k W(x_k - x_{p,j})(u_k - v_j),\label{E:nsdvdt}
\end{align}
where $u_k$ is the velocity of gas in cell $k$, $v_j$ is the velocity of the $j$-th particle, $\epsilon$ is the (uniform) solid-to-gas density ratio, $t_s$ is the stopping time, and $W(x)$ is the particle-mesh weight function.

We now consider the special case of $\epsilon / \tau_s \gg 1$, where $\tau_s$ is the dimensionless stopping time, i.e., $t_s$ normalized by the timescale of interest for the system.
In this limit, equation~\eqref{E:nsdudt} implies that $u_k \approx \sum_j W(x_k - x_{p,j}) v_j$ for all $k$.
We further adopt the CIC weight function such that $W(\pm1/2) = 1/2$ and $W(x) = 0$ for all $|x| \geq \Delta x$.
Equation~\eqref{E:nsdvdt} then becomes
\begin{equation}
\tder{v_j}{t} \approx
\frac{\Delta x^2}{4t_s}
\left(\frac{v_{j+1} - 2v_j + v_{j-1}}{\Delta x^2}\right),\label{E:nsdiffv}
\end{equation}
for all $j$.
This equation is exactly the same as the diffusion equation with a diffusion coefficient of $\Delta x^2 / 4t_s$ when spatially discretized by second-order accurate, centered finite differences.
We note that using a higher-order weight function as TSC does not change this diffusion coefficient, but only increases the accuracy of the spatial derivatives.

Another special case we can consider is the limit of $\tau_s \ll 1$ and $\epsilon / \tau_s \sim 1$.
In this limit, equation~\eqref{E:nsdvdt} implies that $v_j \approx \sum_k W(x_k - x_{p,j}) u_k$ for all $j$.
With the CIC weight function, Equation~\eqref{E:nsdudt} then becomes
\begin{equation}
\tder{u_k}{t} \approx
\frac{\epsilon\Delta x^2}{4t_s}
\left(\frac{u_{k+1} - 2u_k + u_{k-1}}{\Delta x^2}\right),\label{E:nsdiffu}
\end{equation}
for all $k$.
Once again, this equation is the same as the diffusion equation with a diffusion coefficient of $\epsilon\Delta x^2 / 4t_s$, to second-order spatial accuracy.

Equations~\eqref{E:nsdiffv} and~\eqref{E:nsdiffu} indicates that using the particle-mesh method to treat the mutual drag force introduces numerical shear stresses to the system.
Moreover, it can be seen by generalizing the argument outlined above that numerical normal stresses are also induced by the particle-mesh method.
Fortunately, the diffusion coefficient associated with these stresses diminishes quadratically with increasing resolution.
On the other hand, this property implies that the system of equations~\eqref{E:src} is indeed globally coupled, and special care needs to be taken to numerically solve this system both accurately and efficiently.

\section{OPERATOR-SPLITTING METHODS} \label{S:os}

In this appendix, we briefly review the concept of operator splitting, and two standard methods of it.
Suppose that
\begin{equation} \label{E:pdef}
\pder{\vec{\xi}}{t} = (\mathcal{A} + \mathcal{B})\vec{\xi}
\end{equation}
is the full partial differential equation in question, where $\vec{\xi}$ is the vector for the dynamical variables to be solved for and $\mathcal{A}$ and $\mathcal{B}$ are two differential operators.
We can \emph{operator split} equation~\eqref{E:pdef} into two separate differential equations as
\begin{subequations} \label{E:pdes}
\begin{align}
\pder{\vec{\xi}}{t} &= \mathcal{A}\vec{\xi},\\
\pder{\vec{\xi}}{t} &= \mathcal{B}\vec{\xi},
\end{align}
\end{subequations}%
with $\vec{f}(t;\vec{\xi}_0)$ and $\vec{g}(t;\vec{\xi}_0)$ being the respective solutions to equations~\eqref{E:pdes} along with the initial conditions $\vec{\xi}(0) = \vec{\xi}_0$.
Note that $\vec{f}(t;\vec{\xi}_0)$ and $\vec{g}(t;\vec{\xi}_0)$ can be either exact or approximate themselves.
Then by somehow combining $\vec{f}(t;\vec{\xi}_0)$ and $\vec{g}(t;\vec{\xi}_0)$ can the real solution $\vec{\xi}(t)$ to the full equation~\eqref{E:pdef} be approximated. 

There are two standard operator-splitting methods to approximate the solution $\vec{\xi}(t_{n+1})$ at time step $n + 1$ with the initial conditions $\vec{\xi}(t_n) \equiv \vec{\xi}_n$ at time step $n$.
The first is called the Godunov splitting, and $\vec{\xi}(t_{n+1}) \simeq \vec{\xi}_{n+1}$ is given by a two-step method:
\begin{subequations}
\begin{align}
\vec{\xi'} &= \vec{f}(\Delta t; \vec{\xi}_n),\\
\vec{\xi}_{n+1} &= \vec{g}(\Delta t; \vec{\xi'}),
\end{align}
\end{subequations}%
where $\Delta t \equiv t_{n+1} - t_n$ is the size of the time step.
The second is called the Strang splitting \citep{gS68}, and $\vec{\xi}_{n+1}$ is given by a three-step method:
\begin{subequations}
\begin{align}
\vec{\xi'} &= \vec{f}(\frac{1}{2}\Delta t; \vec{\xi}_n),\\
\vec{\xi''} &= \vec{g}(\Delta t; \vec{\xi'}),\\
\vec{\xi}_{n+1} &= \vec{f}(\frac{1}{2}\Delta t; \vec{\xi''}).
\end{align}
\end{subequations}%
The accuracies for the Godunov and the Strang splittings are formally first-order and second-order, respectively.
For a proof of these properties and more information, the interested reader is referred to, e.g., Chapter~17 of \cite{rL02}.


\end{CJK*}

\begin{thebibliography}{}

\bibitem[Alexander et al.(2008)]{AG08}
Alexander, C.~M.~O'D., Grossman, J.~N., Ebel, D.~S., \& Ciesla, F.~J.\ 2008, Science, 320, 1617

\bibitem[ALMA Partnership et al.(2015)]{AB15}
ALMA Partnership, Brogan, C.~L., P{\'e}rez, L.~M., et al.\ 2015, ApJ, 808, L3

\bibitem[Bai \& Stone(2010a)]{BS10a}
Bai, X.-N., \& Stone, J.~M.\ 2010, ApJ, 722, 1437

\bibitem[Bai \& Stone(2010b)]{BS10b}
Bai, X.-N., \& Stone, J.~M.\ 2010, ApJS, 190, 297

\bibitem[Balsara et al.(2009)]{BT09}
Balsara, D.~S., Tilley, D.~A., Rettig, T., \& Brittain, S.~D.\ 2009, MNRAS, 397, 24

\bibitem[Baruteau \& Zhu(2015)]{BZ15}
Baruteau, C., \& Zhu, Z.\ 2015, MNRAS, in press (arXiv:1511.03498)

\bibitem[Benisty et al.(2015)]{BJB15}
Benisty, M., Juhasz, A., Boccaletti, A., et al.\ 2015, A\&A, 578, L6

\bibitem[Booth et al.(2015)]{BSC15}
Booth, R.~A., Sijacki, D., \& Clarke, C.~J.\ 2015, MNRAS, 452, 3932

\bibitem[Brandenburg \& Dobler(2002)]{BD02}
Brandenburg, A., \& Dobler, W.\ 2002, CoPhC, 147, 471

\bibitem[Brandenburg et al.(1995)]{BN95}
Brandenburg, A., Nordlund, A., Stein, R.~F., \& Torkelsson, U.\ 1995, ApJ, 446, 741

\bibitem[Carballido et al.(2006)]{CFP06}
Carballido, A., Fromang, S., \& Papaloizou, J.\ 2006, MNRAS, 373, 1633

\bibitem[Carrera et al.(2015)]{CJD15}
Carrera, D., Johansen, A., \& Davies, M.~B.\ 2015, A\&A, 579, A43

\bibitem[Cuzzi \& Alexander(2006)]{CA06}
Cuzzi, J.~N., \& Alexander, C.~M.~O.\ 2006, Nature, 441, 483

\bibitem[Dipierro et al.(2015)]{DP15}
Dipierro, G., Price, D., Laibe, G., et al.\ 2015, MNRAS, 453, L73

\bibitem[Dong et al.(2015a)]{DZ15}
Dong, R., Zhu, Z., Rafikov, R.~R., \& Stone, J.~M.\ 2015a, ApJ, 809, L5

\bibitem[Dong et al.(2015b)]{DZW15}
Dong, R., Zhu, Z., \& Whitney, B.\ 2015b, ApJ, 809, 93

\bibitem[Epstein(1924)]{pE24}
Epstein, P.~S.\ 1924, PhRv, 23, 710

\bibitem[Fung \& Dong(2015)]{FD15}
Fung, J., \& Dong, R.\ 2015, ApJ, 815, L21

\bibitem[Goldreich \& Lynden-Bell(1965)]{GL65}
Goldreich, P., \& Lynden-Bell, D.\ 1965, MNRAS, 130, 125

\bibitem[Goldreich \& Ward(1973)]{GW73}
Goldreich, P., \& Ward, W.~R.\ 1973, ApJ, 183, 1051

\bibitem[Hawley et al.(1995)]{HGB95}
Hawley, J.~F., Gammie, C.~F., \& Balbus, S.~A.\ 1995, ApJ, 440, 742

\bibitem[Hockney \& Eastwood(1988)]{HE88}
Hockney, R.~W., \& Eastwood, J.~W.\ 1988, Computer Simulation Using Particles (New York, NY: CRC Press)

\bibitem[Inoue \& Inutsuka(2008)]{II08}
Inoue, T., \& Inutsuka, S.-i.\ 2008, ApJ, 687, 303

\bibitem[Jin et al.(2016)]{JL16}
Jin, S., Li, S., Isella, A., Li, H., \& Ji, J.\ 2016, ApJ, in press (arXiv:1601.00358)

\bibitem[Johansen et al.(2014)]{JB14}
Johansen, A., Blum, J., Tanaka, H., et al.\ 2014, in Protostars and Planets VI, ed.~H.~Beuther, R.~S.~Klessen, C.~P.~Dullemond, \& T.~Henning (Tucson, AZ: Univ. Arizona Press), 547

\bibitem[Johansen et al.(2006)]{JHK06}
Johansen, A., Henning, T., \& Klahr, H.\ 2006, ApJ, 643, 1219

\bibitem[Johansen et al.(2015)]{JM15}
Johansen, A., Mac Low, M.-M., Lacerda, P., \& Bizzarro, M.\ 2015, SciA, 1, 1500109

\bibitem[Johansen et al.(2007)]{JO07}
Johansen, A., Oishi, J.~S., Mac Low, M.-M., et al.\ 2007, Nature, 448, 1022

\bibitem[Johansen \& Youdin(2007)]{JY07}
Johansen, A., \& Youdin, A.\ 2007, ApJ, 662, 627

\bibitem[Johansen et al.(2009)]{JYM09}
Johansen, A., Youdin, A., \& Mac Low, M.-M.\ 2009, ApJ, 704, L75

\bibitem[Kokubo \& Ida(1996)]{KI96}
Kokubo, E., \& Ida, S.\ 1996, Icarus, 123, 180

\bibitem[Kowalik et al.(2013)]{KH13}
Kowalik, K., Hanasz, M., W{\'o}lta{\'n}ski, D., \& Gawryszczak, A.\ 2013, MNRAS, 434, 1460

\bibitem[Laibe \& Price(2011)]{LP11}
Laibe, G., \& Price, D.~J.\ 2011, MNRAS, 418, 1491

\bibitem[Laibe \& Price(2012)]{LP12}
Laibe, G., \& Price, D.~J.\ 2012, MNRAS, 420, 2365

\bibitem[LeVeque(2002)]{rL02}
LeVeque, R.~J.\ 2002, Finite Volume Methods for Hyperbolic Problems (Cambridge, United Kingdom: Cambridge Univ.\ Press)

\bibitem[Lor{\'e}n-Aguilar \& Bate(2014)]{LB14}
Lor{\'e}n-Aguilar, P., \& Bate, M.~R.\ 2014, MNRAS, 443, 927

\bibitem[Lor{\'e}n-Aguilar \& Bate(2015)]{LB15}
Lor{\'e}n-Aguilar, P., \& Bate, M.~R.\ 2015, MNRAS, 454, 4114

\bibitem[Lyra et al.(2009)]{LJ09}
Lyra, W., Johansen, A., Klahr, H., \& Piskunov, N.\ 2009, A\&A, 493, 1125

\bibitem[Miniati(2010)]{fM10}
Miniati, F.\ 2010, JCoPh, 229, 3916

\bibitem[Nakagawa et al.(1986)]{NSH86}
Nakagawa, Y., Sekiya, M., \& Hayashi, C.\ 1986, Icarus, 67, 375

\bibitem[Nelson \& Gressel(2010)]{NG10}
Nelson, R.~P., \& Gressel, O.\ 2010, MNRAS, 409, 639

\bibitem[Oishi et al.(2007)]{OMM07}
Oishi, J.~S., Mac Low, M.-M., \& Menou, K.\ 2007, ApJ, 670, 805

\bibitem[Paardekooper \& Mellema(2006)]{PM06}
Paardekooper, S.-J., \& Mellema, G.\ 2006, A\&A, 453, 1129

\bibitem[Safronov(1969)]{vS69}
Safronov, V.~S.\ 1969, Evoliutsiia doplanetnogo oblaka (Moscow: Nauka)

\bibitem[Simon et al.(2015)]{SA15}
Simon, J.~B., Armitage, P.~J., Li, R., \& Youdin, A.~N.\ 2015, ApJ, submitted (arXiv:1512.00009)

\bibitem[Strang(1968)]{gS68}
Strang, G.\ 1968, SJNA, 5, 506

\bibitem[Surville et al.(2016)]{SML16}
Surville, C., Mayer, L., \& Lin, D.~N.~C.\ 2016, ApJ, submitted (arXiv:1601.05945)

\bibitem[Tamura et al.(2009)]{mT09}
Tamura, M., \& SEEDS Team\ 2009, in AIP Conf. Proc.\ 1158, Exoplanets and Disks:~Their Formation and Diversity, ed.~T.~Usuda, M.~Tamura, \& M.~Ishii (Melville, NY: AIP), 11

\bibitem[Tilley et al.(2010)]{TB10}
Tilley, D.~A., Balsara, D.~S., Brittain, S.~D., \& Rettig, T.\ 2010, MNRAS, 403, 211

\bibitem[van der Marel et al.(2013)]{MDB13}
van der Marel, N., van Dishoeck, E.~F., Bruderer, S., et al.\ 2013, Science, 340, 1199

\bibitem[Weidenschilling(1977)]{sW77}
Weidenschilling, S.~J.\ 1977, MNRAS, 180, 57

\bibitem[Whipple(1972)]{fW72}
Whipple, F.~L.\ 1972, in Twenty-First Nobel Symp., From Plasma to Planet, ed.~A.~Evlius (New York, NY: Wiley Interscience Division), 211

\bibitem[Yang \& Johansen(2014)]{YJ14}
Yang, C.-C., \& Johansen, A.\ 2014, ApJ, 792, 86

\bibitem[Yang \& Krumholz(2012)]{YK12}
Yang, C.-C., \& Krumholz, M.\ 2012, ApJ, 758, 48

\bibitem[Youdin(2010)]{aY10}
Youdin, A.~N.\ 2010, EAS Publ.\ Series, 41, 187

\bibitem[Youdin \& Goodman(2005)]{YG05}
Youdin, A.~N., \& Goodman, J.\ 2005, ApJ, 620, 459

\bibitem[Youdin \& Johansen(2007)]{YJ07}
Youdin, A., \& Johansen, A.\ 2007, ApJ, 662, 613

\bibitem[Zhu et al.(2014)]{ZS14}
Zhu, Z., Stone, J.~M., Rafikov, R.~R., \& Bai, X.-N.\ 2014, ApJ, 785, 122

\bibitem[Zsom \& Dullemond(2008)]{ZD08}
Zsom, A., \& Dullemond, C.~P.\ 2008, A\&A, 489, 931

\end{thebibliography}
\end{document}